\documentclass[]{pasj02} 
\usepackage{multirow,url} 

\jyear{2026}
\Received{}
\Accepted{}

\newcommand{\eg}{e.g., }
\newcommand{\ie}{i.e., }
\newcommand{\Msun}{M_{\odot}}

\newcommand{\teff}{$T_{\rm eff}$}
\newcommand{\logg}{$\log (g)$}
\newcommand{\MH}{[M/H]}
\newcommand{\SN}{$S/N$}
\newcommand{\CH}{[C/H]}
\newcommand{\FeH}{[Fe/H]}
\newcommand{\MHnb}{[M/H]$_{\rm NB}$}
\newcommand{\MHnbf}{[M/H]$_{\rm NB,fin}$}
\newcommand{\CHnb}{[C/H]$_{\rm NB}$}
\newcommand{\CHnbf}{[C/H]$_{\rm NB,fin}$}
\newcommand{\CFenb}{[C/Fe]$_{\rm NB}$}

\newcommand{\FeHspec}{[M/H]$_{\rm spec}$}
\newcommand{\FeHhrs}{[Fe/H]$_{\rm HRS}$}
\newcommand{\CHhrs}{[C/H]$_{\rm HRS}$}
\newcommand{\CFehrs}{[C/Fe]$_{\rm HRS}$}
\newcommand{\FeHmalls}{[M/H]$_{\rm MALLS}$}

\newcommand{\Vmic}{$v_{\rm t}$}
\newcommand{\CFe}{[C/Fe]}

\newcommand{\teffm}{$T_{\rm eff,m}$}
\newcommand{\loggm}{$\log (g)_{\rm m}$}
\newcommand{\MHm}{[Fe/H]$_{\rm m}$}
\newcommand{\aFem}{[$\alpha$/Fe]$_{\rm m}$}
\newcommand{\CFem}{[C/Fe]$_{\rm m}$}
\newcommand{\MgFem}{[Mg/Fe]$_{\rm m}$}
\newcommand{\dteff}{$\Delta T_{\rm eff}$}
\newcommand{\dlogg}{$\Delta \log (g)$}
\newcommand{\dMH}{$\Delta$[Fe/H]}
\newcommand{\sMH}{$\sigma_{\rm [M/H]}$}
\newcommand{\sCH}{$\sigma_{\rm [C/H]}$}
\newcommand{\sNB}{$\sigma_{NB395}$}
\newcommand{\sNBup}{$\sigma_{NB395,{\rm up}}$}

\newcommand{\ion}[2]{\text{#1\,\uppercase\expandafter{\romannumeral#2}}}

\begin{document} 

\title{Bright Metal-Poor Star Survey - I. Tomo-e Gozen narrow-band photometric survey and medium-resolution spectroscopic follow-up}

\author{
 Hiroko \textsc{Okada},\altaffilmark{1,2,3}\altemailmark\orcid{0009-0009-6151-8157} \email{hiroko.okada@nao.ac.jp} 
 Takumi \textsc{Iwasaki},\altaffilmark{3}\orcid{} 
 Nozomu \textsc{Tominaga},\altaffilmark{1,4,3}\orcid{0000-0001-8537-3153}
 Wako \textsc{Aoki},\altaffilmark{1,4}\orcid{0000-0002-8975-6829}
 Satoshi \textsc{Honda},\altaffilmark{2}\orcid{0000-0001-6653-8741} 
 Kurumi \textsc{Furutsuka},\altaffilmark{2}\orcid{0009-0008-0515-7492}
 Tadafumi \textsc{Matsuno},\altaffilmark{5}\orcid{0000-0002-8077-4617} 
 Tomoki \textsc{Morokuma},\altaffilmark{6}\orcid{0000-0001-7449-4814} 
 Takuma \textsc{Suda},\altaffilmark{7,8}\orcid{0000-0002-4318-8715} 
 Miho N. \textsc{Ishigaki},\altaffilmark{1,4}\orcid{0000-0003-4656-0241} 
 Naoto \textsc{Kobayashi},\altaffilmark{9}\orcid{0000-0003-4578-2619}
 Hidenori \textsc{Takahashi},\altaffilmark{9}\orcid{} 
 Yuu \textsc{Niino},\altaffilmark{9}\orcid{0000-0001-5322-5076} 
 Satoshi \textsc{Takita},\altaffilmark{10}\orcid{0009-0007-4821-5827} 
 Sohei \textsc{Kondo},\altaffilmark{9}\orcid{} 
 Yuki \textsc{Mori},\altaffilmark{9}\orcid{} 
 Kenzo \textsc{Kinugasa},\altaffilmark{9}\orcid{} 
 Shigeyuki \textsc{Sako},\altaffilmark{10,11,12,13,14}\orcid{0000-0002-8792-2205} 
 Tomio \textsc{Kanzawa},\altaffilmark{1}\orcid{} 
 Hikaru \textsc{Iwashita},\altaffilmark{1}\orcid{} 
 Kenji \textsc{Mitsui},\altaffilmark{1}\orcid{}
 Takeo \textsc{Fukuda},\altaffilmark{1}\orcid{}
 Keiko \textsc{Kaneko},\altaffilmark{1}\orcid{0000-0003-0822-2591} 
 and 
 Mitsuhiro \textsc{Fukushima}\altaffilmark{1}\orcid{} 
}

\altaffiltext{1}{National Astronomical Observatory of Japan, 2-21-1 Osawa, Mitaka, Tokyo 181-8588, Japan}
\altaffiltext{2}{Nishi-Harima Astronomical Observatory, Center for Astronomy, University of Hyogo, 407-2 Nishigaichi, Sayo-cho, Sayo, Hyogo 679-5313, Japan}
\altaffiltext{3}{Department of Physics, Faculty of Science and Engineering, Konan University, 8-9-1 Okamoto, Kobe, Hyogo 658-8501, Japan}
\altaffiltext{4}{Astronomical Science Program, Graduate Institute for Advanced Studies, SOKENDAI, 2-21-1 Osawa, Mitaka, Tokyo 181-8588, Japan}
\altaffiltext{5}{Astronomisches Rechen-Institut, Zentrum für Astronomie der Universität Heidelberg, Mönchhofstraße 12-14, 69120, Heidelberg, Germany}
\altaffiltext{6}{Astronomy Research Center, Chiba Institute of Technology, 2-17-1 Tsudanuma, Narashino, Chiba 275-0016, Japan}
\altaffiltext{7}{Department of Liberal Arts, Tokyo University of
Technology, 5-23-22 Kamata, Ota-ku, Tokyo 144-8535, Japan}
\altaffiltext{8}{Research Center for the Early Universe, The University of Tokyo, 7-3-1 Hongo, Bunkyo-ku, Tokyo 113-0033, Japan}
\altaffiltext{9}{Kiso Observatory, Institute of Astronomy, Graduate School of Science, The University of Tokyo, 10762-30 Mitake, Kiso-machi,
Kiso-gun, Nagano 397-0101, Japan}
\altaffiltext{10}{Institute of Astronomy, Graduate School of Science, The University of Tokyo, 2-21-1 Osawa, Mitaka, Tokyo 181-0015, Japan}
\altaffiltext{11}{UTokyo Organization for Planetary Space Science, The University of Tokyo, 7-3-1 Hongo, Bunkyo-ku, Tokyo 113-0033, Japan}
\altaffiltext{12}{Collaborative Research Organization for Space Science and Technology,
The University of Tokyo, 7-3-1 Hongo, Bunkyo-ku, Tokyo 113-0033, Japan}
\altaffiltext{13}{Research Center for the Early Universe, Graduate School of Science,
The University of Tokyo, 7-3-1 Hongo, Bunkyo-ku, Tokyo
113-0033, Japan}
\altaffiltext{14}{Next-generation Neutrino Science and Multi-messenger Astronomy
Organization, The University of Tokyo, 5-1-5 Kashiwanoha,
Kashiwa, Chiba 277-8582, Japan}




\maketitle

\begin{abstract}
We present the Tomo-e Gozen Bright Metal-Poor Star Survey (TeMPS), a new wide-area narrow-band photometric survey designed to search for bright metal-poor stars in the Northern sky.
The survey uses the Tomo-e Gozen camera on the 1.05 m Kiso Schmidt telescope with four narrow-band filters centered on the \ion{Ca}{2} H and K lines, the CH $G$ band, H$\alpha$, and a reference wavelength region. We review the survey strategy, photometric data processing, and calibration of metallicity and carbon abundance estimates derived from narrow-band colors. We further present medium-resolution spectroscopic follow-up observations with Nayuta/MALLS to validate the photometric selection and identify new metal-poor stars. The current data set covers $\gtrsim 22,000$~deg$^{2}$ in all four bands with a total on-source integration time of $\sim 100$ hr. The median limiting magnitudes of the narrow-band observations, defined at a signal-to-noise ratio of $20$, approximately correspond to $G\sim12.5$, making the survey well suited for constructing bright metal-poor samples for high-resolution spectroscopy. By combining narrow-band photometry with archival broad-band photometry and {\it Gaia} distances, we estimate \teff, \logg, metallicity, and carbon abundance. Calibration against literature abundances derived from high-resolution spectra shows that the photometric method reproduces metallicities and carbon abundances with typical scatters of $\lesssim 0.3$ dex and $\lesssim 0.4$ dex, respectively. We estimate metallicities for $\sim 1.7$ million stars and identify $\sim 16,000$ very metal-poor candidates with \MHnbf $<-2$. We also show that Nayuta/MALLS medium-resolution spectra provide metallicities consistent with those derived from high-resolution spectra, with a scatter of $\sim 0.27$ dex. Among 32 photometrically selected candidates followed up with MALLS, 24 are confirmed to have \FeHmalls $< -2$, including one newly identified star with \FeHmalls $\simeq -3.4$. These results demonstrate that Tomo-e Gozen narrow-band photometry and MALLS medium-resolution spectroscopic follow-up provide an efficient means of selecting bright metal-poor stars for future abundance studies with high-resolution spectroscopy.

\end{abstract}


\section{Introduction}
Metal-poor stars provide a crucial fossil record for investigating the earliest stages of Galactic chemical evolution. In particular, very metal-poor (VMP) stars, commonly defined as stars with [Fe/H]\footnote{Metallicity is characterized by the abundance ratio of metals to H (\MH) and is frequently represented by the abundance ratio of Fe to H (\FeH). Here [A/B] $=\log_{10}(N_{\rm A}/N_{\rm B})-\log_{10}(N_{\rm A}/N_{\rm B})_\odot$, where the subscript $\odot$ refers to the solar value and $N_{\rm A}$ and $N_{\rm B}$ are the abundances of elements A and B, respectively. In this paper, the metallicity derived from high-resolution spectroscopy is denoted by \FeH, while the metallicity derived using other methods is denoted by \MH.} $< -2$, formed at epochs when the metal content of the early Universe was still low. Except for a small number of elements whose surface abundances can be altered by stellar evolutionary processes, the chemical composition observed in their atmospheres preserves that of the natal gas clouds from which they formed. Detailed abundance patterns of such stars can therefore be compared with theoretical nucleosynthesis models, providing constraints on element production in early generations of nuclesynthesis events, e.g., supernovae \citep{nom13}. In addition, observed abundance trends as a function of metallicity encode essential information on the chemical enrichment history of the Milky Way (\eg \cite{fre15}).

Since VMP stars are intrinsically rare, large-scale surveys are required to identify them efficiently. Over the past several decades, extensive spectroscopic and photometric surveys have substantially expanded the known samples of metal-poor stars.
Early searches for metal-poor stars relied on broad-band photometry \citep{rom54,wal60}, high-proper-motion measurements \citep{rya91,car96}, and the strength of the \ion{Ca}{2}~K line \citep{bon70,bid73}. In particular, Schmidt objective-prism spectroscopic surveys targeting the \ion{Ca}{2}~H and K lines near $395$~nm achieved remarkable success, most notably the HK survey \citep{bee92} and the Hamburg/ESO Survey \citep{chr08}. The latter led to the discovery of two hyper metal-poor stars with \FeH~$<-5$, HE~0107--5240 \citep{chr02} and HE~1327--2326 \citep{fre05}. 
Subsequently, low-resolution spectroscopic surveys with multi-object spectrographs, such as the Sloan Extension for Galactic Understanding and Exploration (SEGUE; \cite{segue}) and the Large Sky Area Multi-Object Fiber Spectroscopic Telescope (LAMOST; \cite{zha06}), have greatly increased the number of known metal-poor stars. These surveys have played a central role in determining the metallicity distribution function and in providing metal-poor candidates for large high-resolution spectroscopic follow-up observations \citep{2012A&A...542A..87B,aok13,aok22,li22}. They have also enabled statistical studies of carbon-enhanced metal-poor (CEMP) stars, showing that the CEMP fraction increases toward lower metallicity (e.g., \cite{lee13}). This trend highlights the importance of carbon abundance as an additional diagnostic in searches for the most metal-poor stars.
More recently, photometric surveys employing narrow-band (NB) filters sensitive to the \ion{Ca}{2} H and K lines have become an efficient approach to identifying metal-poor candidates. Pioneering examples include the SkyMapper Southern Sky Survey \citep{kel07} and the Pristine survey \citep{sta17}. The former led to the discovery of SMSS~J031300.36--670839.3, the most iron-poor star known to date, with \FeH~$<-7$ \citep{kel14}, while the latter has provided important insights into the formation and evolution of the Galaxy (\eg \cite{ses20}). Additional narrow-band photometric surveys, including J-PLUS \citep{cen19}, S-PLUS \citep{alm22}, and SAGES \citep{fan23}, have further expanded the search for metal-poor stars.

The availability of {\it Gaia} Data Release 3 (DR3; \cite{GaiaDR3XP1,GaiaDR3XP2,GaiaDR3RV}), including BP/RP (XP) and RVS spectra, has led to numerous catalogue-based attempts to search for metal-poor stars using a variety of methods. For example, XGBoost-based machine-learning methods \citep{and23}, data-driven models \citep{zha23}, tree-based machine-learning models \citep{hat25}, stellar-locus methods \citep{hua25}, and uncertainty-aware cost-sensitive neural networks \citep{yan25} have been applied to {\it Gaia} XP spectra. In addition, Ca triplet equivalent widths have been measured from {\it Gaia} RVS spectra \citep{mat24}, flux ratios have been derived from {\it Gaia} XP spectra \citep{xyl24}, and photometric metallicities have been estimated from {\it Gaia} XP spectra \citep{mar24}. These studies have provided large catalogs of metal-poor star candidates.

Follow-up high-resolution spectroscopic observations are required to confirm the metallicities of metal-poor star candidates identified by the surveys and catalogue-based searches described above and to determine their detailed elemental abundances. The feasibility of such follow-up observations depends strongly on stellar brightness. Bright metal-poor stars allow high signal-to-noise ratio spectra over a broad wavelength range to be obtained with relatively short exposure times, enabling precise abundance measurements for many elements.
Unique insights are obtained when abundances are measured for numerous elements. For example, CS~31082-001, with $G=11.4$ and \FeH~$=-2.8$, has been used to examine the universality of the $r$-process through measurements of about 60 elements \citep{hil02}. HD~122563, with $G=5.9$ and \FeH~$=-2.8$, exhibits a weak $r$-process pattern dominated by first-peak $r$-process elements \citep{hon06}. BD+44$^\circ$493, with $G=8.9$ and \FeH~$=-3.8$, provides evidence for enrichment by faint supernovae \citep{ito09,ito13}. Chemical abundances of SMSS~2003-1142, with $G=11.6$ and \FeH~$=-3.5$, have been measured for 44 elements from C to U, revealing high [Zn/Fe] and enhanced abundances of $r$-process elements, which could be explained by the r-process nucleosynthesis in magneto-rotational supernovae rather than in neutron star mergers \citep{yon21}.

Despite their importance, bright metal-poor stars remain limited in number. This is partly because many previous surveys used to search for metal-poor stars were designed to reach faint magnitudes and were optimized for distant, i.e., high-redshift, galaxies and QSOs as well as metal-poor stars. Consequently, many metal-poor stars identified by such surveys are too faint for detailed high-resolution spectroscopic abundance analyses of many elements. Expanding the sample of bright VMP stars (\eg $G<13$) is therefore essential for increasing the number of objects for which key and rare elements can be measured.

We have therefore undertaken a shallow, wide-area survey of bright metal-poor stars using NB photometry with the Tomo-e Gozen camera. This survey, the Tomo-e Gozen Bright Metal-Poor Star Survey (TeMPS), is designed to discover as many bright metal-poor stars as possible. We use four NB filters centered at 395, 411, 433, and 656~nm to estimate stellar metallicities and carbon abundances. A subset of the metal-poor star candidates is followed up with medium-resolution spectroscopy using Nayuta/MALLS to determine their metallicities. The observing programs with the Tomo-e Gozen camera and Nayuta/MALLS are ongoing. In this paper, we present the first results from the observing campaigns, including the NB photometric survey and spectroscopic follow-up observations, as a proof of concept based on data obtained by 2026. 

This paper is structured as follows. We describe NB photometric surveys in Section~\ref{sec:NBobs} and spectroscopic follow-up observations in Section~\ref{sec:follow-up}. Our results are presented and discussed in Section~\ref{sec:result}. Finally, we present our summary in Section~\ref{sec:summary}. We adopt solar abundances from \citet{2009ARA&A..47..481A} throughout this paper.

\section{Narrow-band photometric survey}
\label{sec:NBobs}

\subsection{Instruments}
\label{sec:instruments}

The Tomo-e Gozen camera (\cite{sak18}; Sako et al. in prep.) is a wide-field camera on the 1.05 m
Kiso Schmidt telescope \citep{tak77}. It consists of 4 quadrant camera units with 21 Complementary Metal Oxide Semiconductor (CMOS)
image sensors. The units are labeled Q1 through Q4. The sensors on the unit are aligned to the spherical focal plane with spatial gaps of
25.0~arcmin in right-ascension and 25.6~arcmin in declination.
Each sensor covers 39.7~arcmin $\times$ 22.4~arcmin. In total, the Tomo-e Gozen camera simultaneously covers 20.8~deg$^2$
with 84 sensors, corresponding to $\sim30\%$ of the entire focal plane
(9$\degree$ in diameter) of the telescope. The wavelength coverage of the CMOS image
sensors ranges from 370~nm to 730~nm (\eg \cite{zha24}; Sako et al. in prep.). Thanks to the
negligible readout time of the CMOS image sensors, \ie a reset loss time of $0.1$~ms per frame, any exposure time
appropriate for individual science cases can be chosen. The 84 sensors have independent windows to which different filters can be applied without any interference.

\begin{figure}
 \begin{center}
  \includegraphics[width=0.8\columnwidth]{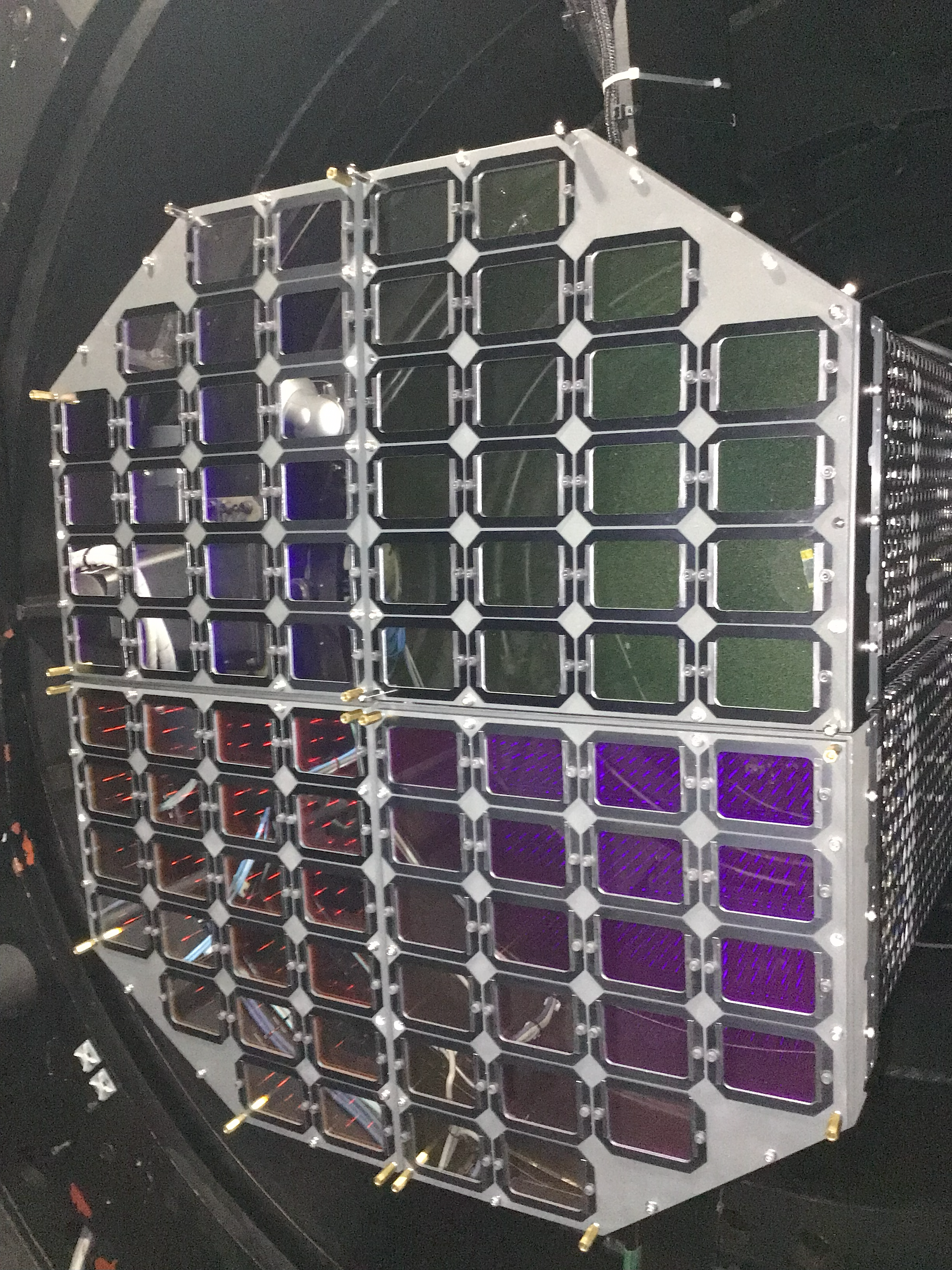} 
 \end{center}
\caption{
Photograph of the
 Tomo-e Gozen camera equipped with 84
 NB filters. The $NB395$, $NB411$, $NB433$, and $NB656$ filters are
 installed in Q1 (right top), Q2 (left top), Q3 (right bottom), and Q4
 (left bottom), respectively.
}\label{fig:filterholder}
\end{figure}

The original window frame of the Tomo-e Gozen camera was flat, and thus the incident light entered the entrance windows at different angles \citep{sak18}. The oblique incidence on the entrance window is not suitable for NB observations because it causes different wavelength shifts of the transmission curves depending on the position of sensors in the focal plane. To mitigate this influence, spherical window frames parallel to the focal plane were installed to the Q1 unit on Jul 2022 and to all units on May 2023. To efficiently produce the curved structures of the window frame and top bracket, they were manufactured using a metal 3-dimensional printer and a CNC milling machine (Sako et al. in prep.). As a result, nearly normal incidence of light on all entrance windows is achieved on average (Figure~\ref{fig:filterholder}).

\subsection{Observations}
\label{sec:survey}

\begin{figure*}
 \begin{center}
  \includegraphics[width=\textwidth]{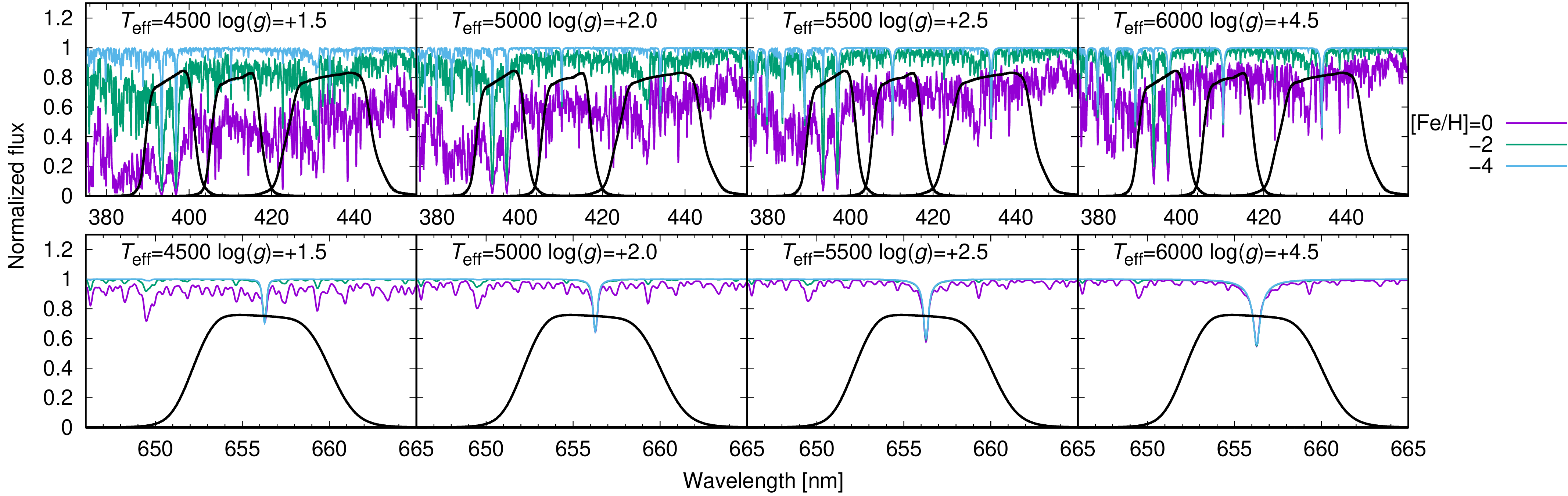} 
 \end{center}
\caption{ Normalized transmission curves of the NB395, NB411, NB433, and
 NB656 filters (black solid lines) compared with normalized  spectra of models with \FeH~$=0$ (purple lines), $-2$ (green lines), and $-4$ (cyan lines). The stellar parameters of model spectra are \teff~$=4500$~K and
 \logg~$=1.5$~dex (left), \teff~$=5000$~K and
 \logg~$=2.0$~dex (middle left), \teff~$=5500$~K and
 \logg~$=2.5$~dex (middle right), and \teff~$=6000$~K and
 \logg~$=4.5$~dex (right).
}\label{fig:filter}
\end{figure*}

For the metal-poor star survey, we designed four NB filters summarized in
Table~\ref{tab:filters} and Figure~\ref{fig:filter}.
The bands centered at 395~nm, 433~nm, and 656~nm cover the
Ca H+K lines, CH $G$-band, and H$\alpha$, respectively, while the band
centered at 411~nm, around which spectral features are relatively sparse, is used as a reference for the NB395 and NB433 bands.
We manufactured 21 filter plates for each narrow band. 

\begin{table}
  \tbl{Properties of NB filters.}{%
  \begin{tabular}{cccc}
   \hline
   Filter name & Effective wavelength & Wavelength width & Feature  \\ 
   & [nm] & FWHM [nm] & \\ 
   \hline
   NB395 & 395.5 & 11.8 & Ca H+K \\
   NB411 & 411.5 & 12.1 & reference \\
   NB433 & 433.9 & 20.9 & CH $G$-band \\
   NB656 & 656.1 & 8.0 & H$\alpha$ \\
   \hline
  \end{tabular}}\label{tab:filters}
\end{table}

\begin{figure*}
 \begin{center}
  \includegraphics[width=\textwidth]{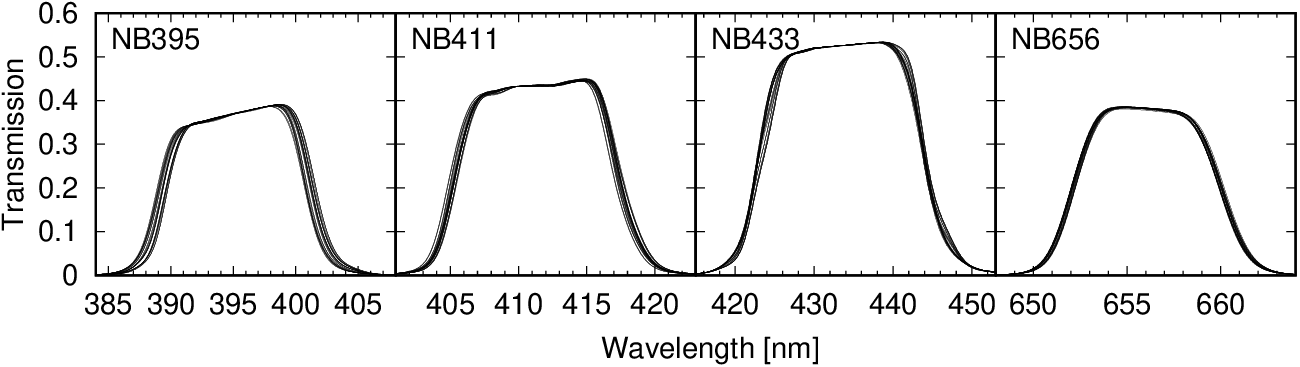} 
 \end{center}
\caption{ Transmission curves of 21 plates of the NB395, NB411, NB433, and
 NB656 filters.
}\label{fig:filter2}
\end{figure*}

The transmission curves of 84 NB filter plates are shown in
Figure~\ref{fig:filter2}. The transmission curves include the
atmospheric extinction assuming the air mass of $1.3$, the
transmission of the Schmidt plate, the reflectivity of mirror, the
widening of transmission curve due to the f$/3.1$
incident beam, and the quantum efficiency of the CMOS image sensors.
Figure~\ref{fig:filter2} demonstrates the fairly homogeneous transmission of 21
plates of each narrow band. The magnitude differences caused by the effects of the small variations among the NB395 and NB433
plates are corrected in the photometry (Section~\ref{sec:photometry}).

We have performed a survey four times using 11 NB395 and 10
NB433 filter plates and the Q1 unit on Sep-Oct, 2022 and using all NB395, NB411, NB433, and NB656
filter plates and all units (84 filter plates in total, Figure~\ref{fig:filterholder}) on May-Jun, 2024, May-Jun, 2025, and Jan-Feb, 2026 (Table~\ref{tab:runs}). The adopted frame rates and the number of frames are summarized in Table~\ref{tab:runs}. While a frame rate of 2~fps was adopted for the first two runs, frame rates of 2~fps and 0.4~fps were adopted for the latter two runs. While the high frame rate is adopted so that bright stars are not saturated, the low frame rate is adopted to reduce the fraction of read noise to the total noise budget because the sky background is relatively low in the NB filter surveys. 

\begin{table*}
  \tbl{Observing runs}{%
  \begin{tabular}{ccccc}
   \hline
   Observing runs & Filters & Frame rate &  Number of frames & Exposure time per cube \\ 
   & & [fps] & & [s]\\ 
   \hline
   \multirow{2}{*}{Sep-Oct 2022} & \multirow{2}{*}{NB395 and NB433} & 2 & 30 & 14.5\\
    & & 2 & 60 & 29.5\\ \hline
   May-Jun 2024 & NB395, NB411, NB433, and NB656 & 2 & 30 &14.5 \\ \hline
   \multirow{2}{*}{May-Jun 2025} & \multirow{2}{*}{NB395, NB411, NB433, and NB656} & 2 & 6 & 2.5 \\
   & & 0.4 & 8 & 17.5\\ \hline
   \multirow{2}{*}{Jan-Feb 2026} & \multirow{2}{*}{NB395, NB411, NB433, and NB656} & 2 & 6 & 2.5\\
   & & 0.4 & 6 & 12.5\\
   \hline
  \end{tabular}
 }\label{tab:runs}
\end{table*}

\begin{figure*}
 \begin{center}
  \includegraphics[width=\textwidth]{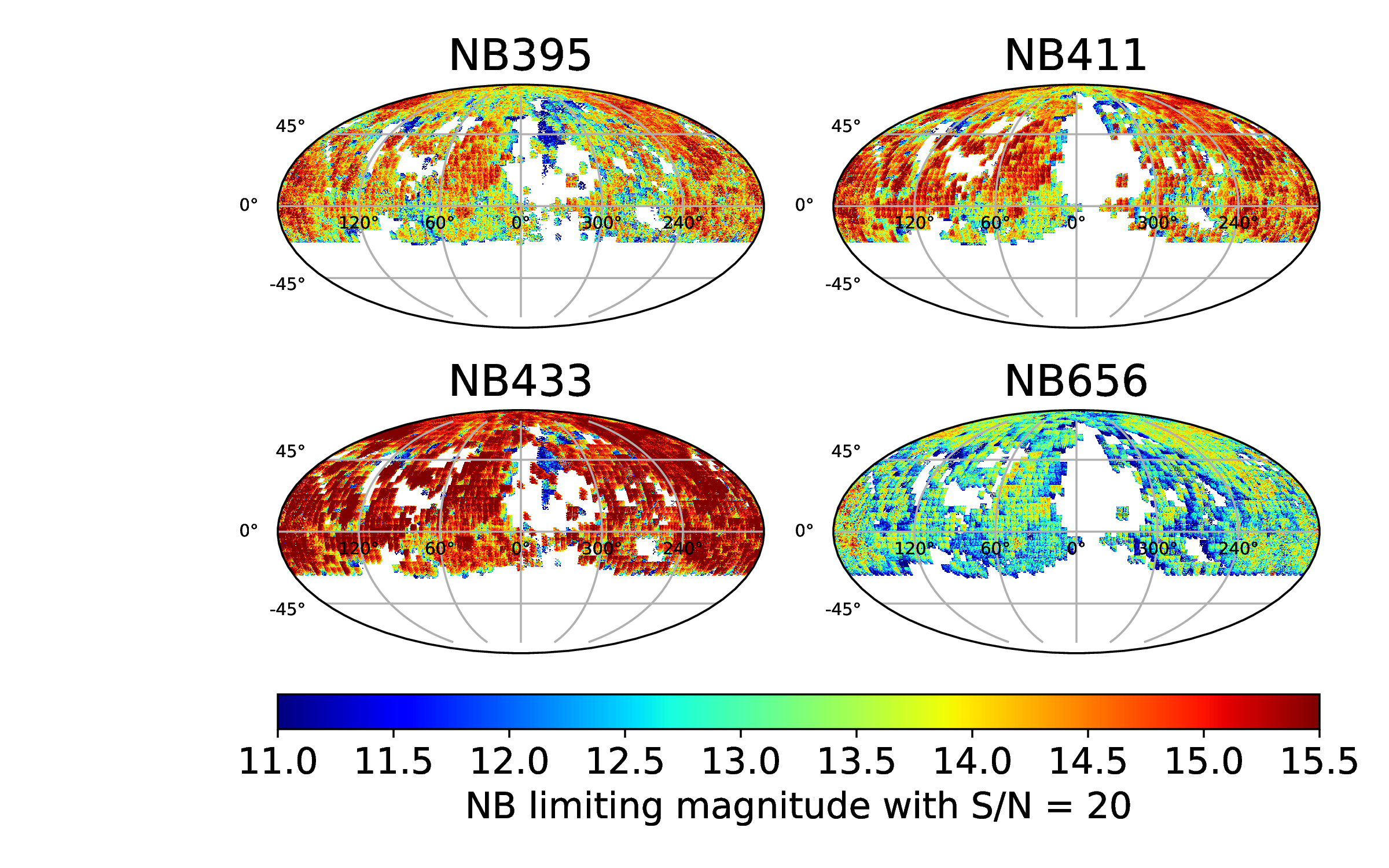} 
 \end{center}
\caption{ Maps of the region surveyed in the $NB395$, $NB411$, $NB433$, and $NB656$ bands. The color represents the NB limiting magnitude at which $S/N=20$ is achieved.
}\label{fig:limmagmap}
\end{figure*}

\begin{figure}
 \begin{center}
  \includegraphics[width=\columnwidth]{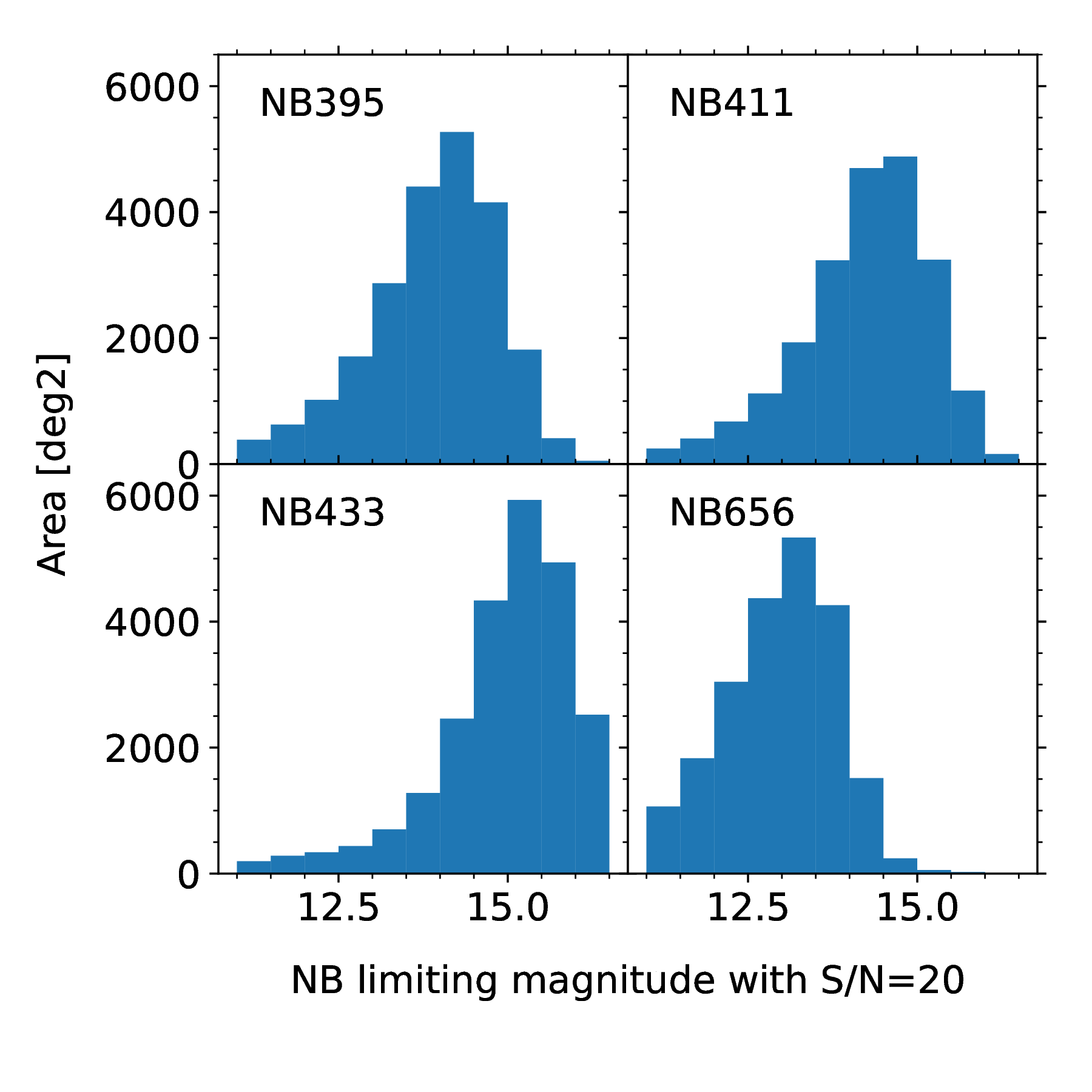} 
 \end{center}
\caption{Survey area as a function of the NB limiting magnitude at which S/N = 20 is achieved for the $NB395$, $NB411$, $NB433$, and
 $NB656$ bands.
}\label{fig:limmaghist}
\end{figure}

The observations have a total on-source integration time of $\sim100$ hours. In the following, only the images, for which astrometry is successfully
solved using \texttt{astrometry.net} \citep{lan10}, are used. 
Figure~\ref{fig:limmagmap} shows the
survey area with the NB limiting magnitude at which the signal-to-noise
ratio (\SN) of 20 is achieved. There are variations in the limiting magnitude due to the
vignetting of the Kiso Schmidt telescope and variations in weather conditions. 
The survey covers $\gtrsim22,000$~deg$^2$ for all filters
(Figure~\ref{fig:limmaghist}).  The median depths
at \SN~$=20$ are $14.0$ in the $NB395$ band,
$14.3$ in the $NB411$ band, $15.2$ in the $NB433$ band, and $13.0$ in the $NB656$ band. The brightness approximately corresponds to $G\sim12.5$. 
The detailed numbers are given in Table~\ref{tab:survey}.

\begin{table}
  \tbl{Details of the survey.}{%
  \begin{tabular}{cccc}
   \hline
   Filter name & Area &  \multicolumn{2}{c}{Median of limiting magnitude} \\ 
   & [deg$^2$] & [NB mag] & [{\it Gaia} $G$ mag] \\ 
   \hline
   NB395 & 23,617 & 14.02 & 12.60 \\
   NB411 & 22,248 & 14.31 & 13.26 \\
   NB433 & 24,855 & 15.16 & 14.18 \\
   NB656 & 23,396 & 13.01 & 13.10 \\
   \hline
  \end{tabular}
 }\label{tab:survey}
\end{table}

\subsection{Data reduction and analysis}
\label{sec:dataanalysis}

\subsubsection{Photometry}
\label{sec:photometry}

The Tomo-e Gozen camera obtains 3-dimensional cube FITS files consisting of multiple frames after the exposure (Sako et al. in prep.). The self-bias subtraction is applied before the generation of cube FITS files. We adopt the standard science reduction procedure to produce a two-dimensional stacked FITS file from the cube FITS file; each frame is corrected using the dark and flat-field frames. The corrected frames in the cube FITS file are averaged after excluding the maximum value, \ie using a clipped mean procedure.

Hot pixels become prominent in NB images because the sky background is relatively low. To exclude hot pixels  from the following analysis, we first co-add 3,000--10,000 individual stacked FITS files obtained in the same observing run and then identify hot pixels on each sensor as those with values exceeding $+10\sigma$ or below $-10\sigma$, where $\sigma$ is the standard deviation of the pixel-value distribution. After masking these hot pixels, we detect sources and perform aperture photometry with a $16.6\arcsec$ ($14$~pixel)-diameter aperture using \texttt{SExtractor} \citep{ber96}. We then solve the astrometry for the stacked FITS files with \texttt{astrometry.net} \citep{lan10}.

\begin{figure*}
 \begin{center}
  \includegraphics[width=0.8\linewidth]{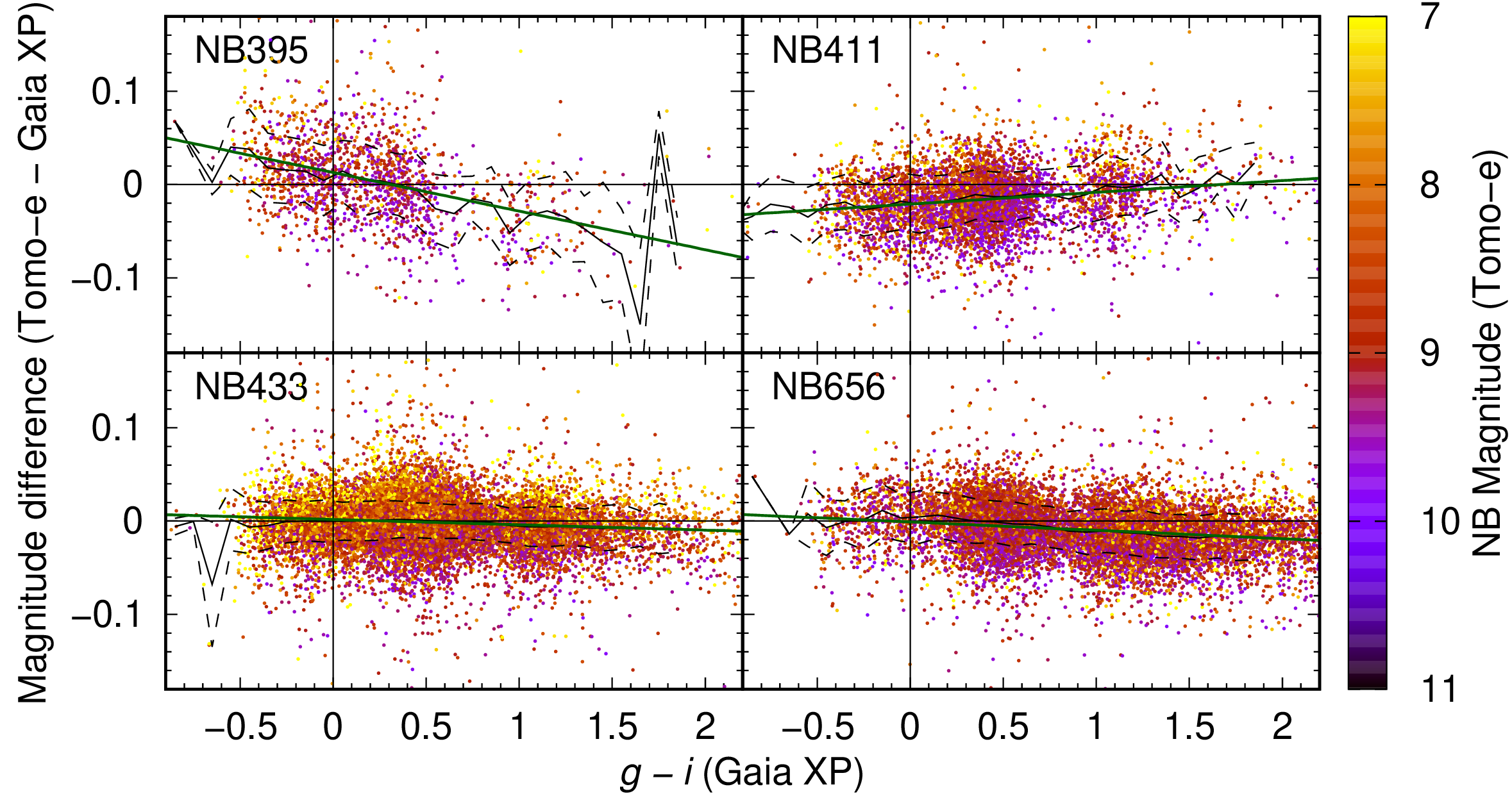} 
 \end{center}
\caption{ Magnitude differences between the $NB395$, $NB411$, $NB433$, and $NB646$ photometry measured with the Tomo-e Gozen camera and those synthesized from the {\it Gaia} XP spectra as a function of $g-i$ color, shown for the reference plates. Point colors indicate the NB magnitude. The adopted color terms are shown by the green lines. The median magnitude difference in each color bin and the corresponding $\pm1\sigma$ scatter are shown by the black solid and dashed lines, respectively.
}\label{fig:colorterm}
\end{figure*}

After the astrometry, sources are detected again with \texttt{SExtractor} and the detected sources are matched with the sources
in the {\it Gaia} DR3 catalog \citep{GaiaDR3,GaiaDR3XP1,GaiaDR3XP2}.
We temporarily estimate the color-independent zero magnitude for each stacked FITS file by comparing the NB magnitudes of
isolated stars\footnote{We define a star with a magnitude $m_{\rm A}$ as isolated if it has no neighboring star with a magnitude $m_{\rm B}$, where $m_{\rm B} < m_{\rm A} + 3$, within a radius of $21\arcsec$.} in the stacked FITS file with those derived for each NB filter plate from the {\it Gaia} XP
spectra. Here, the NB magnitudes are estimated by 
convolving the transmission curve of each filter plate with the
{\it Gaia} DR3 XP spectra derived using \texttt{GaiaXPy}
\citep{GaiaXPy22}.\footnote{https://gaia-dpci.github.io/GaiaXPy-website/}
Even after correcting the color-independent zero magnitude, residual color-dependent magnitude differences remain between the NB photometry with the Tomo-e Gozen camera and those measured with the {\it Gaia} XP spectra (Figure~\ref{fig:colorterm}). For each plate, we derive a color term for each plate with the stars with small photometric errors ($<0.005$~mag). We then apply this term to the NB magnitude derived from the {\it Gaia} XP spectra and determine the color-corrected zero magnitude for each stacked FITS file using the color terms.
Typically, we use $\sim$10 stars for each stacked FITS file to determine the zero magnitude, resulting in a typical uncertainty of $\sim0.05$~mag.

Small plate-to-plate differences in the transmission curves introduce a color dependence in the NB magnitudes derived with different plates. While the color terms for the NB411 and NB656 filter plates are negligible ($<0.005$), those for the NB395 and NB433 plates can be as large as 0.06 and 0.02, respectively, for red stars with $g-i \sim +2$. We therefore derive the plate-to-plate color terms with {\it Gaia} XP spectra and correct the plate-to-plate color terms in the NB395 and NB433 measurements.

For each band, we define a reference plate and derive magnitudes on the system of the corresponding reference NB filter. We compare the colors of targets with those computed using the synthetic spectra and the transmission curve of the reference plate. For targets observed in multiple epochs, we adopt the weighted mean of the individual measurements as the final magnitude. We also estimate the uncertainty of the weighted mean and the epoch-to-epoch scatter as the standard deviation of the measured magnitudes. 
The final magnitude uncertainty is taken to be the geometric mean of these two quantities.

\subsubsection{Estimates of effective temperature and surface gravity}
\label{sec:tefflogg}

Since our targets are bright, their distance and the accurate broad band photometry are
available in the archival catalogues. 
We adopt the photogeometric distance from the {\it Gaia} EDR3
\citep{bai21}. The extinction is corrected using the distance dependent
map of color excess \citep{gre18} and the extinction curve of the Milky Way
\citep{pei92}.

Using $V_{\rm T}$-band magnitudes from the Tycho-2 Catalogue
\citep{tycho} or $V$-band magnitudes from the AAVSO Photometric All-Sky Survey (APASS) DR10\footnote{\url{https://www.aavso.org/apass}} when $V_{\rm T}$ is unavailable, together with $K_s$-band magnitudes from Two Micron All Sky Survey
(2MASS, \cite{tmass}) and correcting for Galactic reddening, we determine the effective temperatures \teff\ and bolometric fluxes of targets are determined using the color-\teff\ relations calibrated with the infrared flux method for dwarf
stars \citep{cas10} and giant stars \citep{alo99}. The surface gravity, \logg, of each target is obtained using its bolometric flux, distance, and effective temperature, assuming a mass of $0.8\Msun$.
We here assume the low metallicity \FeH~$=-3$ as our targets are VMP stars because the metallicity is not estimated in advance. 
Adopting \FeH~$=-1.5$ instead would shift the estimated \teff\ and \logg\ by typically $-140$ to $+30$~K and $-0.10$ to $+0.04$~dex, respectively.

\begin{figure*}
 \begin{center}
  \includegraphics[width=0.7\textwidth]{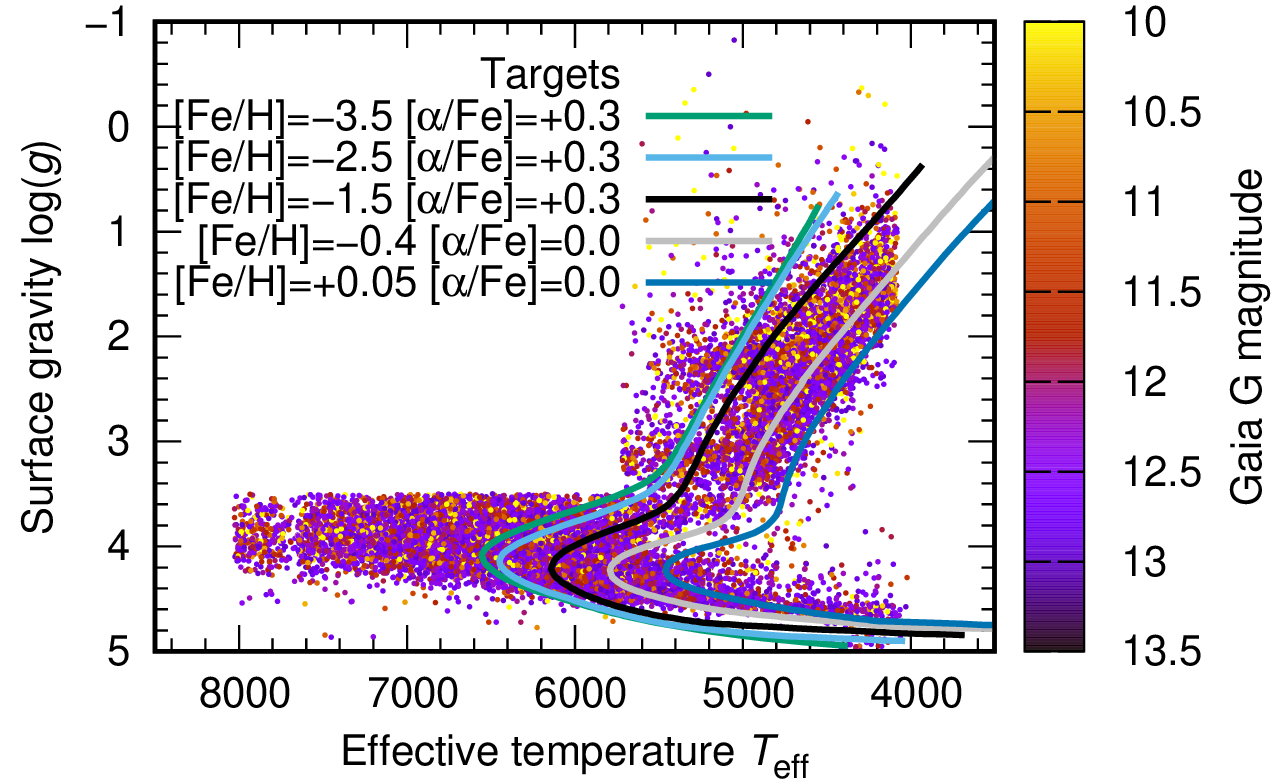} 
 \end{center}
\caption{ \teff-\logg\ plane. The \teff\ and \logg\ values of the targets are shown
 as colored dots. The color represents the apparent $G$ magnitude. Isochrones for an age of 14 Gyr are also shown for \FeH~$=-3.5$ and [$\alpha$/Fe]~$=+0.3$
 (green line), \FeH~$=-2.5$ and [$\alpha$/Fe]~$=+0.3$
 (cyan line), \FeH~$=-1.5$ and [$\alpha$/Fe]~$=+0.3$
 (black line), \FeH~$=-0.4$ and [$\alpha$/Fe]~$=0.0$
 (gray line), and \FeH~$=+0.05$ and [$\alpha$/Fe]~$=0.0$
 (blue line) \citep{dem04}.
}\label{fig:hr}
\end{figure*}

Figure~\ref{fig:hr} shows
the distribution of the targets in the \teff-\logg\ plane and compares
them with theoretical isochrones at the age of $14$~G years \citep{dem04}. The \teff\ and
\logg\ values of a fraction of the targets are roughly consistent with these isochrones, suggesting that they may be old stars, while the other stars with \teff~$>6500$~K are likely to be young stars with \FeH~$>-1$. 

\subsubsection{Estimate of metallicity}
\label{sec:MH}

We construct a grid of synthetic spectra using model atmospheres with various effective
temperature \teffm, surface gravity \loggm, and metallicity
\MHm\ with \texttt{Turbospectrum} \citep{alv98,ple12} in the package of \texttt{iSpec} \citep{2014A&A...569A.111B}\footnote{\url{https://www.blancocuaresma.com/s/iSpec}}. Here, the subscript m denotes the parameters adopted for the synthesis of model spectra.
We assume local thermodynamic equilibrium (LTE) and use one-dimensional hydrostatic \texttt{MARCS} model atmospheres \citep{2008A&A...486..951G}. Atomic line data are primarily taken from the Vienna Atomic Line Database (VALD; \cite{2011BaltA..20..503K}). 

We adopt the
ranges of \teffm\ and \loggm\ referring
\teff\ and \logg\ of observed stars (Figure~\ref{fig:hr}), and the range of \MHm\ from
$-5$ to $+1$. The steps are 100~K for \teffm, $0.1$~dex for \loggm, and
$0.1$~dex for \MHm. 
The abundances of elements are
scaled from the solar abundances with the metallicity, except for those
of $\alpha$-elements \aFem, which gradually increase at lower
metallicities. We set \aFem\ constant at $+0.4$~dex and $0$~dex
in the models with \MHm\ $\leq-1.0$ and $\geq0.0$, respectively, and adopt \aFem~$=-0.4\times$~\MHm\ for the models with $-1.0\leq$ \MHm\ $\leq0.0$ (\eg \cite{mcw97}). 

The magnitudes of model spectra are
derived by convolving the transmission curves of the NB
filters (Section~\ref{sec:survey}) and SDSS filters \citep{doi10}. The 
$g-i$, $NB395-g$, $NB395-g-1.5(g-i)$, $NB433-g$, $NB395-NB411$, $NB395-NB433$,
$NB411-NB433$, $NB656-r$, $NB656-z$, and $NB395-NB656-2(g-i)$ colors are derived for comparisons with those of targets and the determination of the properties of targets. Since
available colors depend on targets, we adopt multiple combinations of
colors as summarized in Table~\ref{tab:colorComb}.

\begin{table*}
  \tbl{Combination of colors to estimate the metallcity and carbon abundance.}{%
  \begin{tabular}{cl}
   \hline
   Name & Colors  \\ 
   \hline
   A & $g-i$, $NB395-g$ \\
   B & $g-i$, $NB395-g-1.5(g-i)$ \\
   C & $g-i$, $NB395-g-1.5(g-i)$, $NB395-NB411$ \\
   D & $g-i$, $NB395-g-1.5(g-i)$, $NB395-NB411$, $NB411-NB433$ \\
   E & $g-i$, $NB395-g-1.5(g-i)$, $NB395-NB411$, $NB433-g$ \\
   F & $g-i$, $NB395-g-1.5(g-i)$, $NB395-NB411$, $NB411-NB433$,
       $NB656-r$ \\
   G & $g-i$, $NB395-g-1.5(g-i)$, $NB395-NB411$, $NB433-g$, $NB656-r$ \\
   H & $g-i$, $NB395-g-1.5(g-i)$, $NB395-NB433$ \\
   I & $g-i$, $NB395-g-1.5(g-i)$, $NB395-NB433$, $NB656-z$ \\
   J & $g-i$, $NB395-g-1.5(g-i)$, $NB395-NB433$, $NB395-NB656-2(g-i)$ \\
   K & $g-i$, $NB395-g-1.5(g-i)$, $NB395-NB433$, $NB656-z$, $NB395-NB656-2(g-i)$ \\
   L & $g-i$, $NB395-g-1.5(g-i)$, $NB395-NB411$, $NB411-NB433$,
       $NB656-z$ \\
   M & $g-i$, $NB395-g-1.5(g-i)$, $NB395-NB411$, $NB411-NB433$,
       $NB395-NB656-2(g-i)$ \\
   N & $g-i$, $NB395-g-1.5(g-i)$, $NB395-NB411$, $NB411-NB433$,
       $NB656-z$, $NB395-NB656-2(g-i)$ \\
   \hline
  \end{tabular}}\label{tab:colorComb}
\end{table*}

In the subsequent analysis, to reduce the bogus classification and avoid missing metal-poor stars in the selection, we adopt relaxed quality cuts as follows:
\begin{enumerate}
 \item Low probability of a source to be variable, $\log_{10}\left(g_{\rm var}\right)<-1.8$, where $g_{\rm var}$ is defined by Equation (3) in \citet{mar24}
 \item Good astrometry, {\it Gaia} Renormalized Unit Weight Error ({\tt RUWE})~$<1.4$
 \item Good {\it Gaia} photometry, $\left|C^*\right|<3\times\sigma_C^*$, where $C^*$ and $\sigma_C^*$ are defined in Equations (6) and (18) in \citet{rie21}
 \item {\it Gaia} {\tt non\_single\_star} flag is $0$
 \item {\it Gaia} {\tt phot\_variable\_flag} is not ``VARIABLE''
\end{enumerate} 

The colors of a target are compared with the colors derived from model spectra within the ranges of \teff\ difference, \dteff, and \logg\ difference, \dlogg, around \teff\ and \logg\ of the target, which are estimated in Section~\ref{sec:tefflogg}, respectively.
The metallicity is estimated only for the targets having the colors inside the range of the colors of compared models.
The metallicity of a target is set to be \MHm\ of the nearest model, \ie with the smallest
distance. The distance in color space $D$ is calculated with 
\begin{equation}
     D = \sqrt{\sum_c \left(S_c - M_c\right)^2},
\end{equation}
where $c$ is the color, $S_c$ and $M_c$ are the colors of a star and a
model spectrum, respectively.

Because line strengths depend on \teff\ and \logg, adopting narrower ranges in \teff\ and \logg\ generally yields more accurate estimates. 
However, narrower parameter ranges imply narrower color ranges, which reduces the number of targets for which metallicity can be estimated. We therefore need to balance this trade-off.
To pick up satisfactory sets of the \teff\ and \logg\ ranges and color combinations in terms of accuracy and number of targets,
we test various ranges of \dteff\ and \dlogg, and the upper limit on the error of NB395 photometry \sNBup. Here, we adopt the ranges of \dteff\ 
$=25$, $50$, $100$, $250$, $500$~K, and infinite, \dlogg\ 
$=0.05$, $0.1$, $0.3$, $0.5$~dex, and infinite, and \sNBup\ $=0.02$, $0.03$, $0.04$, $0.05$, $0.1$, $0.2$, and $0.5$~dex. 

We use the targets observed in 2022 and 2024 to identify satisfactory sets of \teff\ and \logg\ ranges and color combinations by comparing the metallicities estimated using each set with the metallicities derived from high-resolution spectra \FeHhrs. Here, we adopt the measurements of \FeHhrs\ obtained by the Apache Point
Observatory Galactic Evolution Experiment data release 17 (APOGEE DR17, \cite{apogee,apogeedr17}) and the
Galactic Archaeology with HERMES survey data release 4 (GALAH
DR4, \cite{galah4}), or
compiled by the Stellar Abundances for the Galactic Archeology
(SAGA) Database ver. 20230410 \citep{sud08}. The metallicity of targets \MHnb\ is derived for each set of the ranges and color combination, with an offset correction applied.

The satisfactory sets of the \teff\ and \logg\ ranges and color combinations are selected with criteria as follows; (1) The number of applicable targets with
the metallicity measurement with high-resolution spectroscopy is $100$
or more. (2) The number of stars with \FeHhrs~$<-1.5$ is $10$ or more. (3)
A true positive rate, \ie the ratio of targets estimated to have
\MHnb~$<-1.5$ to targets with \FeHhrs~$<-1.5$, is higher than $0.5$. (4) A
false positive rate, \ie the ratio of targets estimated to have
\MHnb~$<-1.5$ to targets with \FeHhrs~$>-1.5$, is less than $0.005$. (5) A
standard deviation of difference between \MHnb\ and \FeHhrs\ is
less than $0.3$. Here, we adopt a threshold metallicity of \FeH~$=-1.5$ to ensure a sufficient number of targets in the true-positive regime. 
The number of satisfactory sets of \teff\ and \logg\ ranges and color combinations is $\sim58,000$ out of $\sim6,500,000$ sets of ranges and color combinations. We average \MHnb\ estimated using the above satisfactory sets and determine the final metallicity of targets \MHnbf.

\begin{figure*}
 \begin{center}
  \includegraphics[width=\textwidth]{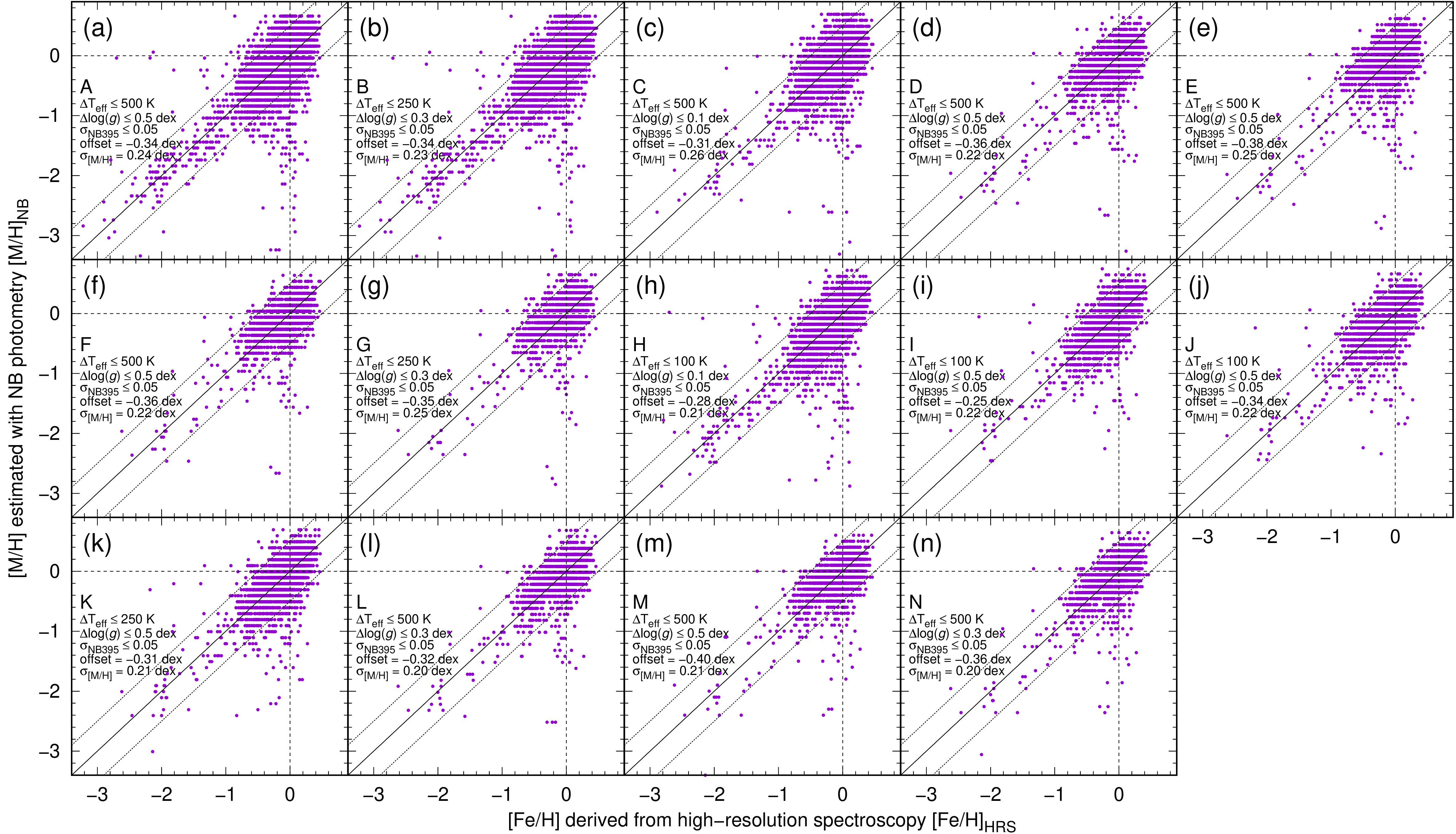}
 \end{center}
\caption{Comparison between the metallicity estimated with the NB photometry \MHnb\ and the metallicity derived from the high-resolution spectroscopy \FeHhrs. Panels (a)-(n) show examples of the results obtained with color combinations A-N, respectively. 
}\label{fig:MHest}
\end{figure*}

Figures~\ref{fig:MHest}(a)-(n) show comparisons between \FeHhrs\ and \MHnb\ derived
with the satisfactory sets of color combinations A-N, respectively. The adopted \dteff, \dlogg, and \sNBup\ are shown in each panel and summarized in Table~\ref{tab:MHest}. The applied
offset and standard deviation are also shown. The offsets
typically range from $-0.3$ to $-0.4$~dex. The offsets stem from the fact that the $NB395-g-1.5(g-i)$ color of a model spectrum is bluer than that of a target at a given $g-i$ color and metallicity. The standard deviations are as
small as $\sim0.2$~dex. The values of offsets do not significantly change even if they are estimated only from the targets with \FeHhrs~$<-1.5$. Typical changes of offsets are less than $0.1$~dex. The metallicities of most targets are
well determined and are consistent with \FeHhrs\ within $\pm0.5$~dex.

\begin{table*}
  \tbl{Examples of \MHnb\ estimates shown in Figures~\ref{fig:MHest}(a)-(n)}{%
  \begin{tabular}{cccccccc}
   \hline
   Combination & Number of stars & \dteff & \dlogg & 
		   \sNB & offset & \sMH & Figure \\ 
   & & [K] & [dex] & & [dex] & [dex] &  \\ 
   \hline
   A & $9241$ & $500$ & $0.5$ & $0.05$ & $-0.34$ & $0.24$ & Figure~\ref{fig:MHest}(a) \\
   B & $8677$ & $250$ & $0.3$ & $0.05$ & $-0.34$ & $0.23$ & Figure~\ref{fig:MHest}(b) \\
   C & $4761$ & $500$ & $0.1$ & $0.05$ & $-0.31$ & $0.26$ & Figure~\ref{fig:MHest}(c) \\
   D & $3147$ & $500$ & $0.5$ & $0.05$ & $-0.36$ & $0.22$ & Figure~\ref{fig:MHest}(d) \\
   E & $3147$ & $500$ & $0.5$ & $0.05$ & $-0.38$ & $0.25$ & Figure~\ref{fig:MHest}(e) \\
   F & $2571$ & $500$ & $0.5$ & $0.05$ & $-0.36$ & $0.22$ & Figure~\ref{fig:MHest}(f) \\
   G & $2194$ & $250$ & $0.3$ & $0.05$ & $-0.35$ & $0.25$ & Figure~\ref{fig:MHest}(g) \\
   H & $5989$ & $100$ & $0.1$ & $0.05$ & $-0.28$ & $0.21$ & Figure~\ref{fig:MHest}(h) \\
   I & $3497$ & $100$ & $0.5$ & $0.05$ & $-0.25$ & $0.22$ & Figure~\ref{fig:MHest}(i) \\
   J & $3497$ & $100$ & $0.5$ & $0.05$ & $-0.34$ & $0.22$ & Figure~\ref{fig:MHest}(j) \\
   K & $4020$ & $250$ & $0.5$ & $0.05$ & $-0.31$ & $0.21$ & Figure~\ref{fig:MHest}(k) \\
   L & $2562$ & $500$ & $0.3$ & $0.05$ & $-0.32$ & $0.20$ & Figure~\ref{fig:MHest}(l) \\
   M & $2615$ & $500$ & $0.5$ & $0.05$ & $-0.40$ & $0.21$ & Figure~\ref{fig:MHest}(m) \\
   N & $2562$ & $500$ & $0.3$ & $0.05$ & $-0.36$ & $0.20$ & Figure~\ref{fig:MHest}(n) \\
   \hline
  \end{tabular}}\label{tab:MHest}
\end{table*}

\begin{figure*}
 \begin{center}
  \includegraphics[width=0.8\textwidth]{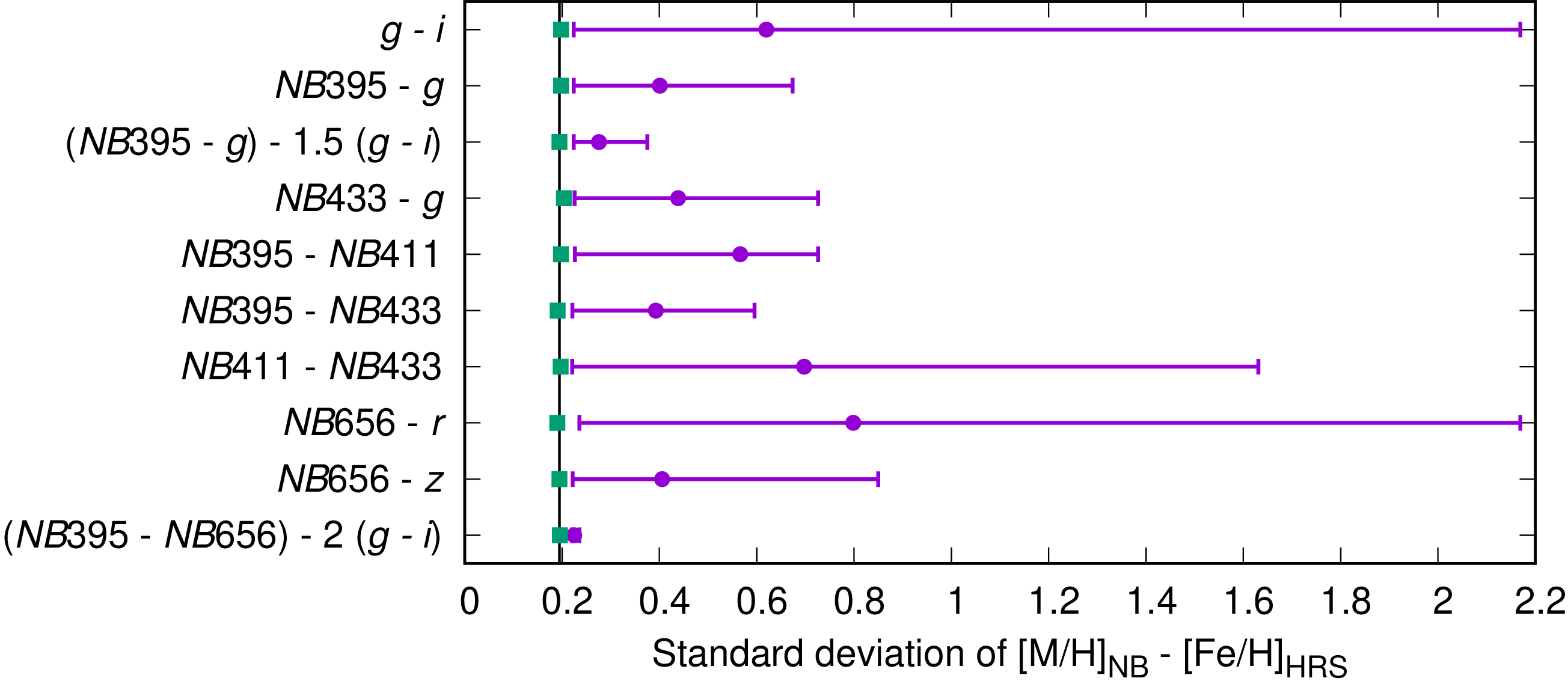}
 \end{center}
\caption{Standard deviations of \MHnb~$-$~\FeHhrs\ for individual colors. The minimum, mean, and maximum standard deviations for the two-color combinations that include each color are shown in magenta, and the standard deviation for the nine-color combinations that exclude that color is shown in green. The standard deviation of the ten-color combination is shown as a black vertical line for guidance. The analysis adopts \dteff~$=500$~K, \dlogg~$=0.3$~dex, and \sNBup~$=0.05$~dex.
}
\label{fig:ColorImportance}
\end{figure*}

We examine the contribution of each of the ten colors to the metallicity estimation by comparing the standard deviations of \MHnb~$-$~\FeHhrs\ for two cases: (i) metallicities derived from two-color combinations selected from the ten colors, and (ii) metallicities derived from nine-color combinations obtained by excluding one color. 
We fix the ranges to \dteff~$=500$~K, \dlogg~$=0.3$~dex, and \sNBup~$=0.05$~dex.
Figure~\ref{fig:ColorImportance} shows, for each color, the minimum, mean, and maximum standard deviation among the two-color combinations that include that color, together with the standard deviation obtained from the nine-color combinations that excludes that color. 
The nine-color combinations yield a standard deviation comparable to the ten-color combination regardless of which color is removed, implying that no single color is essential when the remaining nine colors are available. 
In contrast, the results of two-color combinations depend on the color selection: the smallest standard deviation is obtained with $NB395-NB656-2(g-i)$, followed by $NB395-g-1.5(g-i)$. 
The effectiveness of these two colors is demonstrated in the color-color diagrams (Section~\ref{sec:2colors}). 
The $g-i$, $NB411-NB433$, and $NB656-r$ colors show comparatively larger scatter, and excluding them does not degrade the precision of \MHnb\ estimate.

\subsubsection{Estimate of carbon abundance}
\label{sec:CH}

To estimate the carbon abundance, we construct synthetic spectra of
models with various \teffm, \loggm, \MHm, carbon abundance \CFem, and magnesium
abundance \MgFem, as in the previous
section. For the CH molecular transitions, we adopt the line lists from \citet{MPV}.\footnote{A modified version on Mar 16, 2017 is adopted.} The adopted ranges of \teffm\ and \loggm\ are the same as in
section~\ref{sec:MH}, but the steps are 200~K for \teffm\ and 0.2~dex for
\loggm. The ranges of \MHm, \CFem, and \MgFem\ are $-5$ to $+1$, $-1$ to
$+4$, and $-1$ to $+3$, respectively.
The same colors and quality cuts are adopted as in section~\ref{sec:MH}.

The colors of a target are compared with the colors of model spectra within the ranges of \dteff, \dlogg, and \FeH\ difference, \dMH, around \teff, \logg, and \MHnb\ of the star, respectively.
We adopt the same ranges of \dteff\ and \dlogg, and \sNBup\ as in section~\ref{sec:MH}, and the range of \dMH\ $=0.1, 0.2, 0.5,$ and infinite. The stellar metallicities, \MHnb, are derived in section~\ref{sec:MH}. 
As in section~\ref{sec:MH}, carbon abundance of a target is set to be \CFem\ of the
nearest model.

To identify the satisfactory sets of the ranges and color combinations, we compare the estimated carbon abundance relative to H for the targets observed in 2022 and 2024 with those derived with
high-resolution spectroscopy \CHhrs. After applying the offset, we derive the carbon abundance of the targets, \CHnb,\footnote{Here, we define the carbon abundance as \CHnb~$=$~\CFenb~$+$~\MHnb.} for each set. The final carbon abundance \CHnbf\ is estimated as the average of \CHnb\ estimated with the satisfactory sets.

\begin{figure*}
 \begin{center}
  \includegraphics[width=0.7\textwidth]{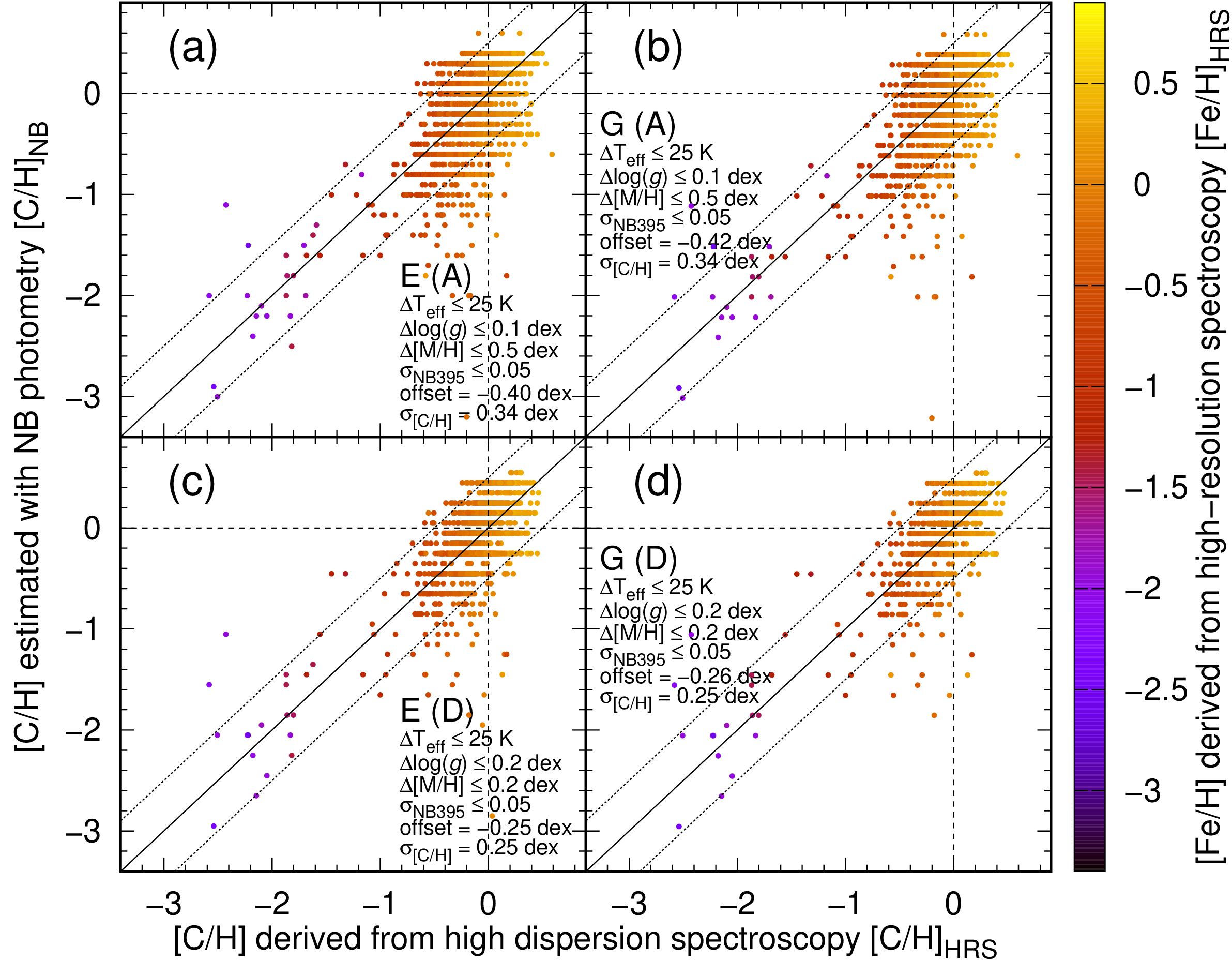}
 \end{center}
\caption{Comparison between the carbon abundance estimated with the NB photometry \CHnb\ and the carbon abundance derived from the high-resolution spectroscopy \CHhrs. The panels show examples of the results obtained with (a) combination E and \MHnb\ estimates from 
 combination set A, (b) combination G and \MHnb\ estimates from combination set A, (c) combination E and \MHnb\ estimates from combination set D, and (d) combination G and \MHnb\ estimates from combination set D.
}\label{fig:MHestC}
\end{figure*}

Figures~\ref{fig:MHestC}(a)-(d) show comparisons between \CHhrs\ and \CHnb\ derived with the combinations E and G with the adopted parameter ranges and resulting offsets summarized in Table~\ref{tab:CHest}. The metallicity
is derived with the combinations A and D shown in Figure~\ref{fig:MHest}(a) and \ref{fig:MHest}(d), respectively. 
The offsets typically range from $-0.2$ to $-0.4$~dex. In contrast to the metallicity estimate, the narrow ranges of \dteff\ and \logg\ are required. This is probably because the strengths of the carbon molecular lines are sensitive to the effective temperature. The standard
deviations can be as small as $\sim0.2-0.4$~dex. This demonstrates that carbon abundances can be well determined from NB photometry alone if the $NB433$ band is included.

\begin{table*}
  \tbl{Examples of \CHnb\ estimates shown in Figures~\ref{fig:MHestC}(a)-(d)}{%
  \begin{tabular}{cccccccccc}
   \hline
   Combination & Combination & Number of stars & \dteff & \dlogg & \dMH & 
		   \sNB & offset & \sCH & Figure \\ 
   for [C/H] & for [M/H] & & [K] & [dex] & [dex] & & [dex] & [dex] &  \\ 
   \hline
   E & A & $2124$ & $25$ & $0.1$ & $0.5$ & $0.05$ & $-0.40$ & $0.34$ & Figure~\ref{fig:MHestC}(a) \\
   G & A & $1758$ & $25$ & $0.1$  & $0.5$ & $0.05$ & $-0.42$ & $0.34$ & Figure~\ref{fig:MHestC}(b) \\
   E & D & $1548$ & $25$ & $0.2$  & $0.2$ & $0.05$ & $-0.25$ & $0.25$ & Figure~\ref{fig:MHestC}(c) \\
   G & D & $1277$ & $50$ & $0.1$  & $0.5$ & $0.05$ & $-0.26$ & $0.25$ & Figure~\ref{fig:MHestC}(d) \\
   \hline
  \end{tabular}}\label{tab:CHest}
\end{table*}

\section{Medium-resolution spectroscopic follow-up}\label{sec:follow-up}

\subsection{Observations}
\label{sec:MALLSobs}
\begin{figure*}
 \begin{center}
  \includegraphics[width=0.7\textwidth]{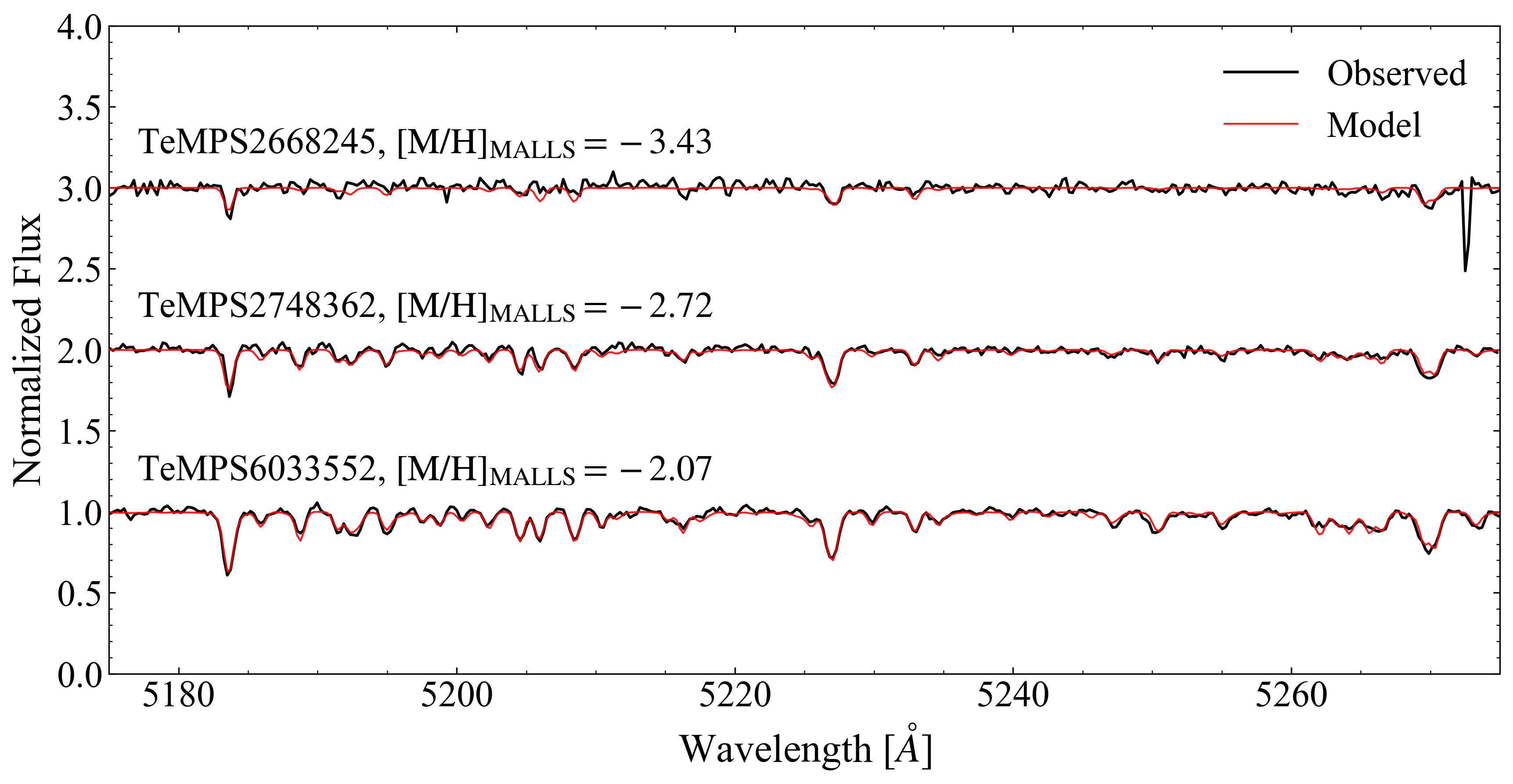}
 \end{center}
\caption{Examples of reduced MALLS spectra for three metal-poor star candidates with different metallicities. The black and red lines show the observed spectra and the best-fit synthetic spectra, respectively. The spectra are vertically offset for clarity. The inferred metallicities \FeHmalls\ are shown for each object.
}
\label{fig:MALLSspec}
\end{figure*}

Spectroscopic observations were carried out with the Nayuta 2.0~m telescope at the Nishi-Harima Astronomical Observatory, University of Hyogo. 
We used the Medium And Low-dispersion Long-slit Spectrograph (MALLS; \cite{oza05}) with the 1800\,lines\,mm$^{-1}$ grating and a $1.2\arcsec$ slit, providing a resolving power of $R \sim 7500$. 
The wavelength coverage was set to $4900$--$5300$\,\AA, simultaneously covering the \ion{Mg}{1} triplet and nearby metal lines.
We observed 27 stars with \FeHhrs\ and 32 metal-poor star candidates identified in Section~\ref{sec:NBobs}. 
Exposure times were chosen to achieve $S/N > 30$ at $\sim5100$\,\AA. 
To mitigate the impact of cosmic rays, individual integrations were limited to a maximum of 1200~s, and the number of exposures was increased depending on the seeing conditions and cloud coverage.
Examples of the MALLS spectra are shown in Figure~\ref{fig:MALLSspec}. These illustrate the quality of the reduced spectra and the metallicity dependence of the absorption features.

\subsection{Data reduction and metallicity measurement}
\label{sec:MALLSreduction}

Data reduction was performed in the standard manner using the  {\tt IRAF}\footnote{IRAF is distributed by the National Optical Astronomy Observatories, which is operated by the Association of Universities for Research in Astronomy, Inc. under cooperative agreement with the National Science Foundation.} packages \texttt{twodspec} and \texttt{onedspec}. 
The reduction procedure included overscan correction, dark subtraction, flat fielding, aperture determination, spectral extraction, wavelength calibration, and continuum normalization. 

Continuum normalization was performed using the \texttt{continuum} task in {\tt IRAF}. 
We first derived normalized spectra by fitting the continuum while excluding pixels with fluxes more than one standard deviation of the continuum flux below the fitted continuum level. 
If the resulting metallicity was higher than or equal to $-1.5$, we adopted that value. If the resulting metallicity was lower than $-1.5$, we reprocessed the spectra using a more conservative rejection threshold of two standard deviations below the fitted continuum level and re-estimated the metallicity.

We estimate the stellar metallicity, \FeHmalls, from the medium-resolution spectra taken with Nayuta/MALLS using a synthetic spectral-fitting method in \texttt{iSpec} because spectral lines are frequently unresolved at medium resolution. Here, we employ \texttt{Turbospectrum} \citep{alv98,ple12} for spectral synthesis as in Sections~\ref{sec:MH} and \ref{sec:CH}. In \texttt{iSpec}, the best fit model is selected to minimize the difference between the observed spectrum and a synthetic spectrum with the Levenberg-Marquardt least-squares minimization \citep{2014A&A...569A.111B}. 

In the synthetic spectrum fitting, we adopt \teff\ and \logg\ estimated in Section~\ref{sec:tefflogg}. The microturbulent velocity \Vmic\ is calculated with an empirical relation (\eg \cite{smi14,jor14}). We vary only the metallicity \MHm\ and the $\alpha$ abundance \aFem\ and identify the model spectra that best fit the medium-resolution spectra. 
Figure~\ref{fig:MALLSspec} shows best-fit synthetic spectra along with the spectra of three metal-poor star candidates with different metallicities. This demonstrates that the spectral fitting reproduces the spectral features well over a wide metallicity range.

\subsection{Metallicity comparisons}
\label{sec:MALLSvalidation}
\begin{figure}[htbp]
  \begin{center}
    \includegraphics[clip,width=0.9\columnwidth]{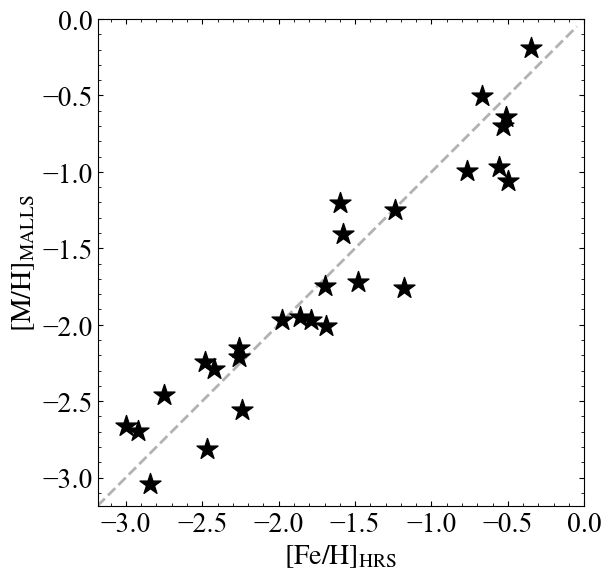}
  \end{center}
    \caption{Comparison between the metallicity determined from the medium-resolution spectroscopy with Nayuta/MALLS, \FeHmalls, and that derived from the high-resolution spectroscopy, \FeHhrs.
}
    \label{fig:MRHR}
\end{figure}

To examine the reliability of the metallicity estimates from the MALLS spectra, we compare \FeHmalls\ with \FeHhrs\ values taken from the literature. Figure~\ref{fig:MRHR} shows the comparison between \FeHhrs\ and \FeHmalls\ for 27 stars whose abundances were measured from high-resolution spectra, spanning a wide metallicity range of $-3 \lesssim \mathrm{[Fe/H]_{HRS}} \lesssim 0$.

The metallicities derived with MALLS spectra are broadly consistent with those derived with high-resolution spectra over the full metallicity range shown in Figure~\ref{fig:MRHR}, indicating that the MALLS spectra provide reliable metallicity estimates at least for the purpose of confirming metal-poor candidates. The agreement is particularly important in the metal-poor regime, \FeHhrs~$\lesssim -2$, where the stars lie close to the one-to-one line with no obvious systematic deviation. The residuals, defined as \FeHmalls~$-$~\FeHhrs, have a mean value of $-0.057$~dex and a standard deviation of $0.27$~dex. The small mean offset indicates that there is no significant global bias relative to the measurements with high-resolution spectra, while the standard deviation represents the typical uncertainty of the metallicities derived with MALLS spectra.
The typical uncertainty is likely due to the spectral features being affected by line blending, continuum placement, or calibration uncertainties. Nevertheless, the overall agreement demonstrates that metallicities derived with MALLS spectra are sufficiently accurate for efficiently confirming metal-poor candidates selected from the photometric survey.

\section{Discussion}
\label{sec:result}

\subsection{Color-color diagrams}
\label{sec:2colors}

Previous studies utilize color-color diagrams to pick up metal-poor star candidates (\eg \cite{SMSSDR1.1,pla22}).
In this section, we plot the colors of targets that have measurements from high-resolution spectroscopy and satisfy the quality cuts. The colors of targets are derived with the NB photometry using the Tomo-e Gozen camera and the broad-band photometry using Gaia XP spectra. 

\begin{figure*}
 \begin{center}
  \includegraphics[width=\textwidth]{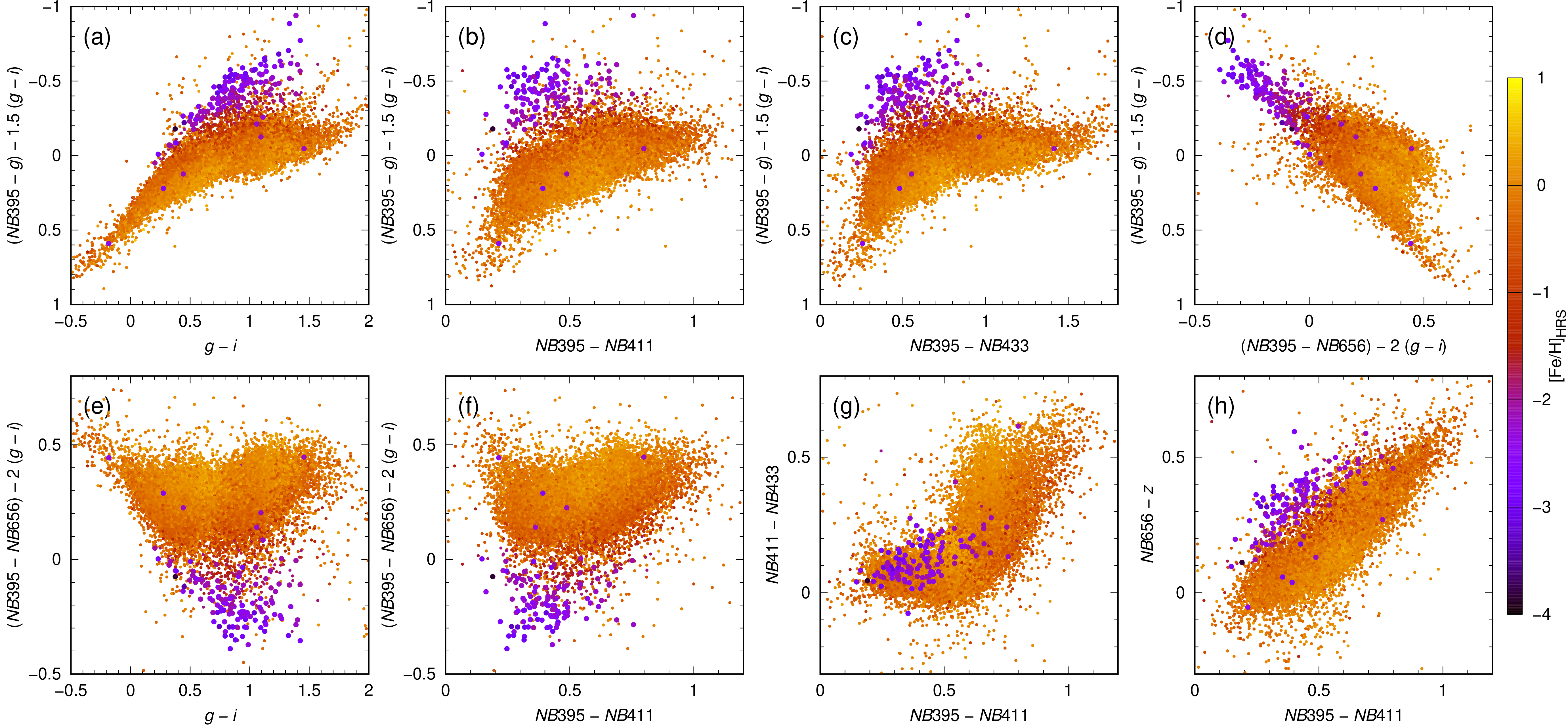}
 \end{center}
\caption{ Color-color diagrams of targets with metallicities derived from high-resolution spectroscopy, \FeHhrs. The color represents \FeHhrs. The panels show diagrams using (a) $g-i$ and
 $NB395-g-1.5(g-i)$, (b) $NB395-NB411$ and
 $NB395-g-1.5(g-i)$, (c) $NB395-NB433$ and
 $NB395-g-1.5(g-i)$, (d) $NB395-NB656-2(g-i)$ and
 $NB395-g-1.5(g-i)$, (e) $g-i$
 and $NB395-NB656-2(g-i)$, (f) $NB395-NB411$
 and $NB395-NB656-2(g-i)$, (g) $NB395-NB411$
 and $NB411-NB433$, and (h) $NB395-NB411$
 and $NB656-z$. Stars with \FeHhrs~$< -2$ are shown with larger symbols for clarity.
 }\label{fig:Color}
\end{figure*}

Figures~\ref{fig:Color}(a)-(h) show the colors and metallicity of targets. We
adopt the colors $g-i$, $NB395-g-1.5(g-i)$, $NB395-NB411$, $NB411-NB433$, $NB395-NB433$, $NB395-NB656-2(g-i)$, and $NB656-z$, which are used in the metallicity estimates in Section~\ref{sec:MH}. The metal-poor stars form clearly separated clusters in some color-color diagrams, although some stars with solar metallicity overlap with them. As shown in previous studies, the metal-poor stars are well separated in the $NB395-g-1.5(g-i)$ color (Figure~\ref{fig:Color}(a)-(d)) and the $NB395-NB656-2(g-i)$ color (Figure~\ref{fig:Color}(d)-(f)). 
At a fixed $g-i$ color, the typical separations in the $NB395-g-1.5(g-i)$ and $NB395-NB656-2(g-i)$ colors between [Fe/H]~$=0$ and $-2$ are $\sim0.3$~mag and $\sim0.4$~mag, respectively.

In addition to the NB395 and NB656 filters, the survey utilizes the NB411 and NB433 filters. These colors are useful for distinguishing metal-poor stars when combined with $NB395-g-1.5(g-i)$ and $NB656-z$, as shown in Figures~\ref{fig:Color}(b), (c), and (h).
The metal-poor stars tend to have blue $NB395-NB411$ and $NB411-NB433$ colors. In particular, Figure~\ref{fig:Color}(b) illustrates that the combination of $NB395-g-1.5(g-i)$ and $NB395-NB411$ clearly separates metal-poor stars from stars with solar metallicity. At a fixed $NB395-NB411$ color, the typical separation in the $NB395-g-1.5(g-i)$ color is $\sim0.4$~mag, being wider than that at a fixed $g-i$ color. 
While we adopt the model fitting to estimate the metallicity in this paper, these color-color diagrams can be used to identify the metal-poor star candidates.

\begin{figure*}
 \begin{center}
  \includegraphics[width=\textwidth]{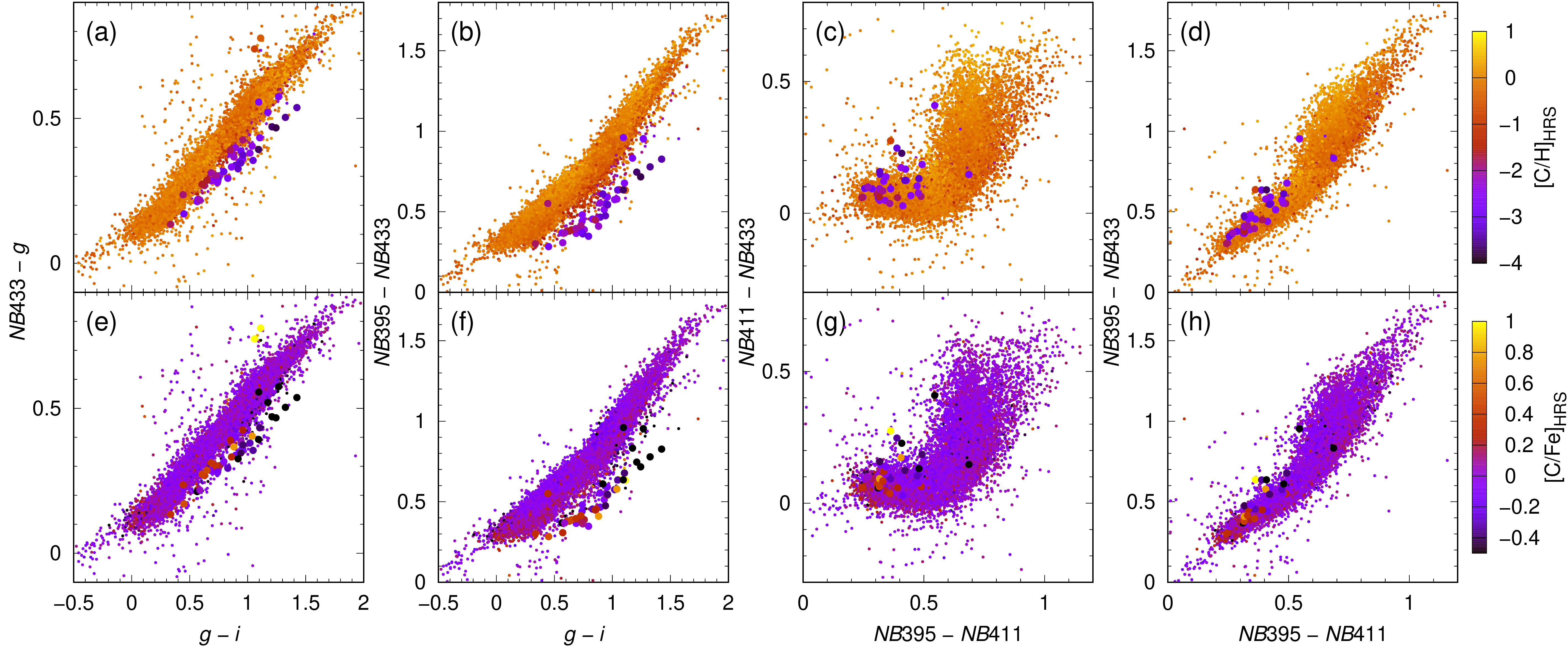}
 \end{center}
\caption{ Color-color diagrams of targets with carbon abundance derived from high-resolution spectroscopy, \CHhrs\ and \CFehrs. The color represents \CHhrs\ (top) and \CFehrs\ (bottom). The panels show diagrams using (a) $g-i$ and
 $NB433-g$, colored by \CHhrs, (b) $g-i$ and
 $NB395-NB433$, colored by \CHhrs, (c) 
 $NB395-NB411$ and
 $NB411-NB433$, colored by \CHhrs, (d) 
 $NB395-NB411$ and
 $NB395-NB433$, colored by \CHhrs, (e) $g-i$ and
 $NB433-g$, colored by \CFehrs, (f) $g-i$ and
 $NB395-NB433$, colored by \CFehrs, (g) 
 $NB395-NB411$ and
 $NB411-NB433$, colored by \CFehrs, and (h) 
 $NB395-NB411$ and
 $NB395-NB433$, colored by \CFehrs. 
 Stars with \FeHhrs~$< -2$ are shown with larger symbols for clarity.
 }\label{fig:ColorC}
\end{figure*}

Figures~\ref{fig:ColorC}(a)-(h) show the colors and carbon abundance of
stars. We adopt the colors $g-i$, $NB433-g$, $NB395-NB433$, $NB395-NB411$, and $NB411-NB433$ as in Section~\ref{sec:CH}. While the separations in the diagrams using $g-i$ and $NB433-g$ or $NB395-NB411$ and $NB411-NB433$ are small, the combination of the $g-i$ and 
$NB395-NB433$ colors is a good indicator of carbon abundance (Figure~\ref{fig:ColorC}(b)). Moreover, since the strengths of CH molecular lines depend directly on the \CH\ ratio, it is better to directly estimate the \CH\ ratio instead of the \CFe\ ratio. The variation in the $NB395-NB433$ color between [C/H]~$=0$ and $-3$ is as large as $\sim0.3$~mag at a given $g-i$ color. This demonstrates that the color-color diagram can be used to estimate the carbon abundance.

\subsection{Comparisons with other studies}

We applied the methods for estimating metallicity and carbon abundance to the targets with $G<13$. Among $\sim4,800,000$ stars with $G<13$ in the northern sky (Dec~$>-30^\circ$), $\sim2,800,000$ targets have NB395 photometric measurements, and we estimate metallicities for $\sim1,700,000$ targets. The number of VMP star candidates with \MHnbf~$<-2$ is $\sim16,000$.

\begin{figure*}
 \begin{center}
  \includegraphics[width=0.7\textwidth]{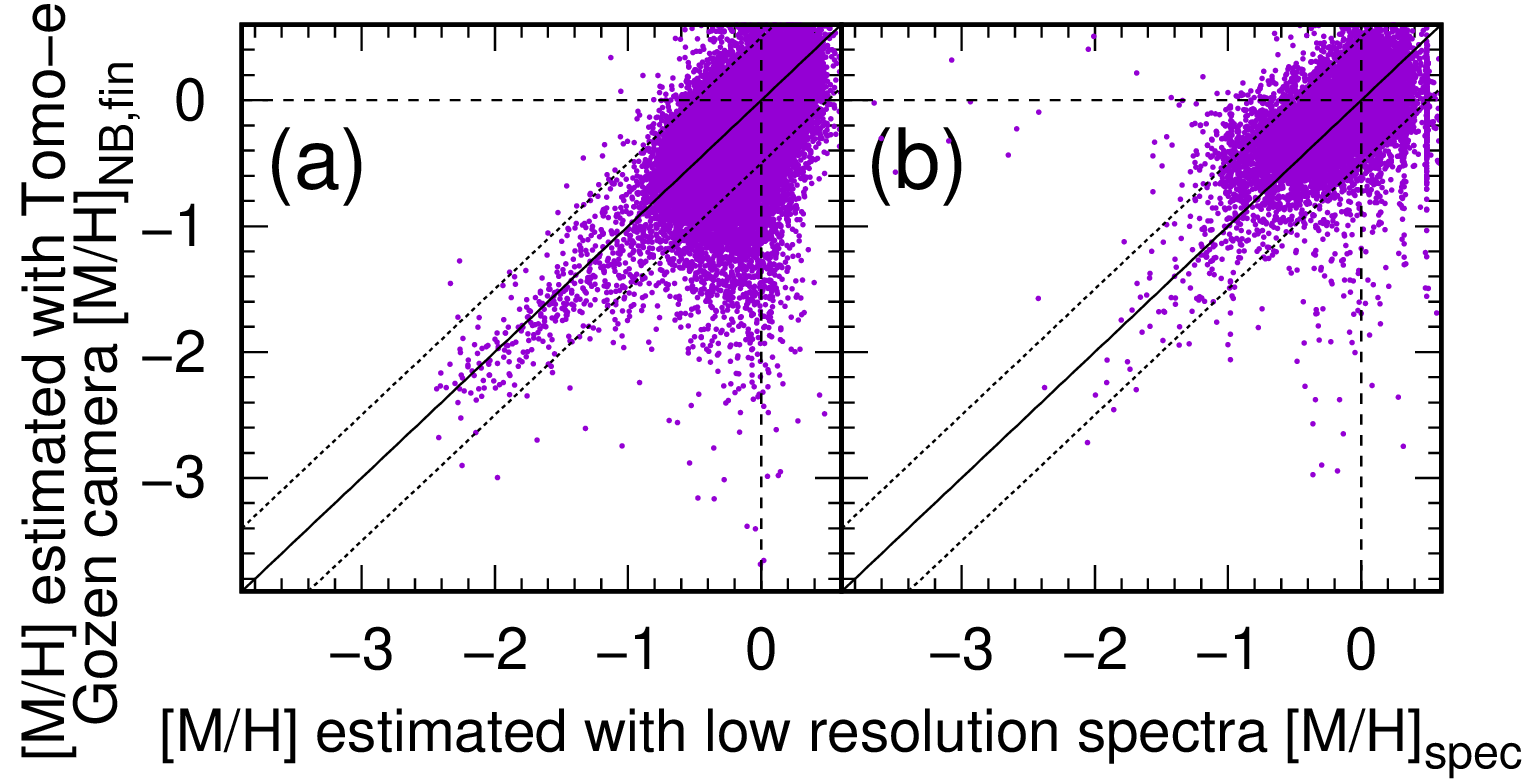}
 \end{center}
\caption{Comparison between the metallicity estimated from NB photometry obtained with the Tomo-e
 Gozen camera, \MHnbf, and that determined from low-resolution spectra, \FeHspec: (a) LAMOST DR10 LRS \citep{lamost} and (b) RAVE DR6 \citep{rave}. 
}\label{fig:CompLAMOST}
\end{figure*}

Our \MHnbf\ estimates are compared with the metallicity measurements, \FeHspec, derived from low-resolution spectra in LAMOST DR10 \citep{lamost} and the Radial Velocity Experiment data release 6 (RAVE DR6 \cite{rave}) (Figure~\ref{fig:CompLAMOST}). Although metallicity estimates are also available from the Sloan Extension for Galactic Understanding and Exploration (SEGUE,
\cite{segue}), there is little overlap in the magnitude ranges.
The \MHnbf\ estimates are roughly consistent with those derived using low-resolution spectra, but there are discrepancies for some stars.  

While few targets with \MHnbf~$>-2$ exhibit \FeHspec~$<-2$ in the RAVE DR6 and LAMOST DR10 catalogues, there are targets with \MHnbf~$<-2$ and \FeHspec~$>-2$ in the comparisons. The targets with false measurements typically have \sNB~$>0.03$, and their stellar parameters are clustered at \teff~$\sim4500-5000$~K, \logg~$\sim+2-+2.5$~dex, and $g-i\sim+1$. This is because, at a given {\it Gaia} $G$ magnitude, giants are faint in the $NB395$ band, and thus the resultant \SN\ is low. This problem can be mitigated by adding more exposures and increasing the \SN.

\begin{figure*}
 \begin{center}
  \includegraphics[width=\textwidth]{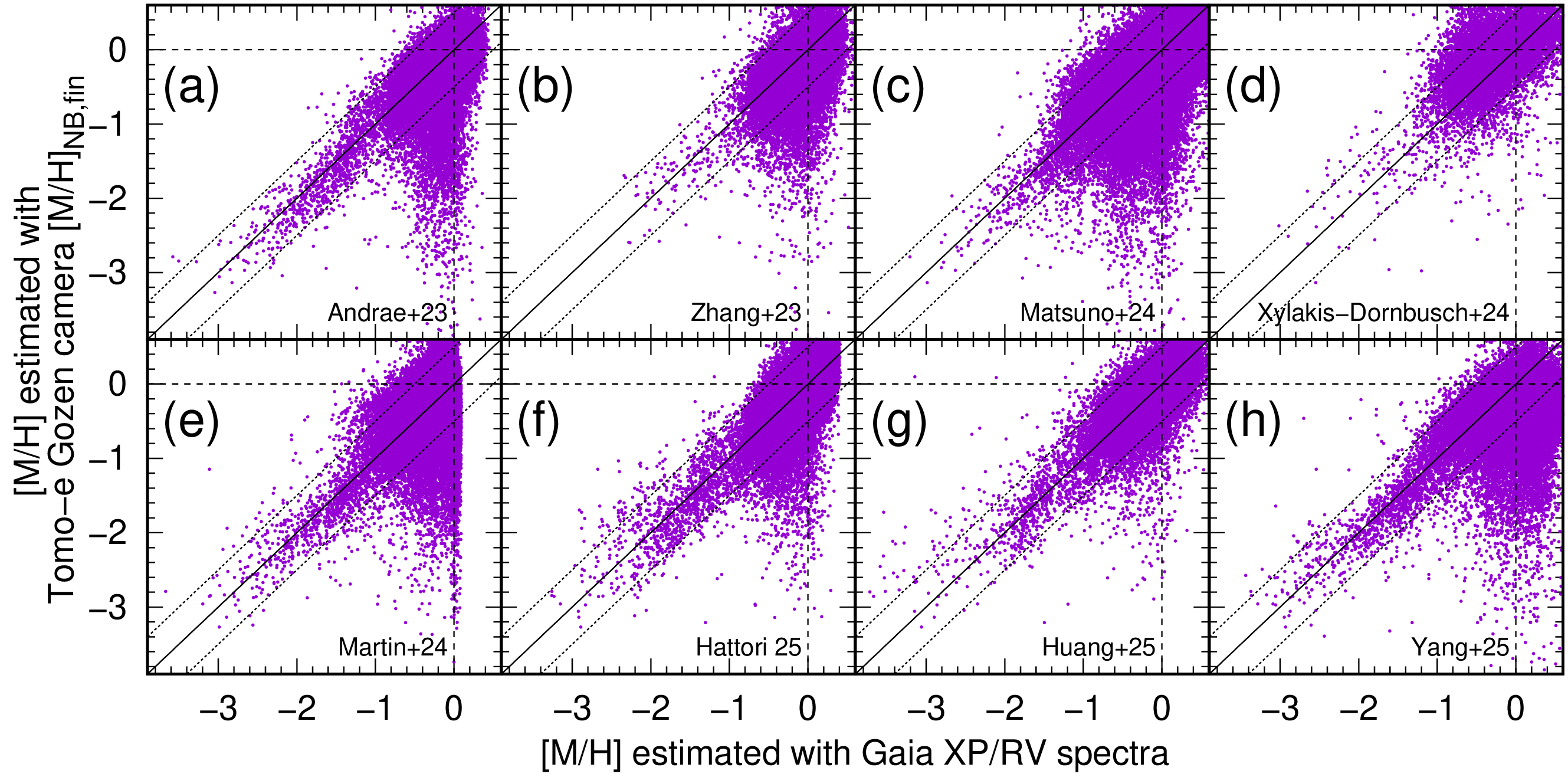}
 \end{center}
\caption{Comparison between the metallicity estimated from NB photometry obtained with the Tomo-e
 Gozen camera, \MHnbf, and those reported in the following 
 studies: (a) \citet{and23}, (b) \citet{zha23}, (c) \citet{mat24}, (d)
 \citet{xyl24}, (e) \citet{mar24}, (f) \citet{hat25}, (g) \citet{hua25}, and
 (h) \citet{yan25}. 
}\label{fig:Comp}
\end{figure*}

Catalogues of metallicities estimated from the {\it Gaia} XP or RVS spectra are available \citep{and23,zha23,mat24,xyl24,mar24,hat25,hua25,yan25}. We compare our metallicity estimates for the targets with these results, adopting the quality cuts used in each study (Figures~\ref{fig:Comp}(a)-(h)). Here we adopt the targets without any spectroscopic metallicity measurements in APOGEE DR17 \citep{apogeedr17},
GALAH DR10 \citep{galah4}, LAMOST DR10 \citep{lamost}, SEGUE \citep{segue}, and RAVE DR6 \citep{rave}. Although there is significant scatter, the \MHnbf\ estimates show good agreement with the metallicity estimates in these studies.
This diagram shows that few targets have \MHnbf~$>-2$ and \MH~$<-2$ derived with other studies, whereas there are targets at \MHnbf~$<-2$ and \MH~$>-2$ derived with other studies. This implies that TeMPS achieves high completeness at the expense of purity, as expected from our strategy of identifying the metal-poor stars using a combination of NB photometry and medium-resolution spectroscopy.

\begin{figure*}
 \begin{center}
  \includegraphics[width=\textwidth]{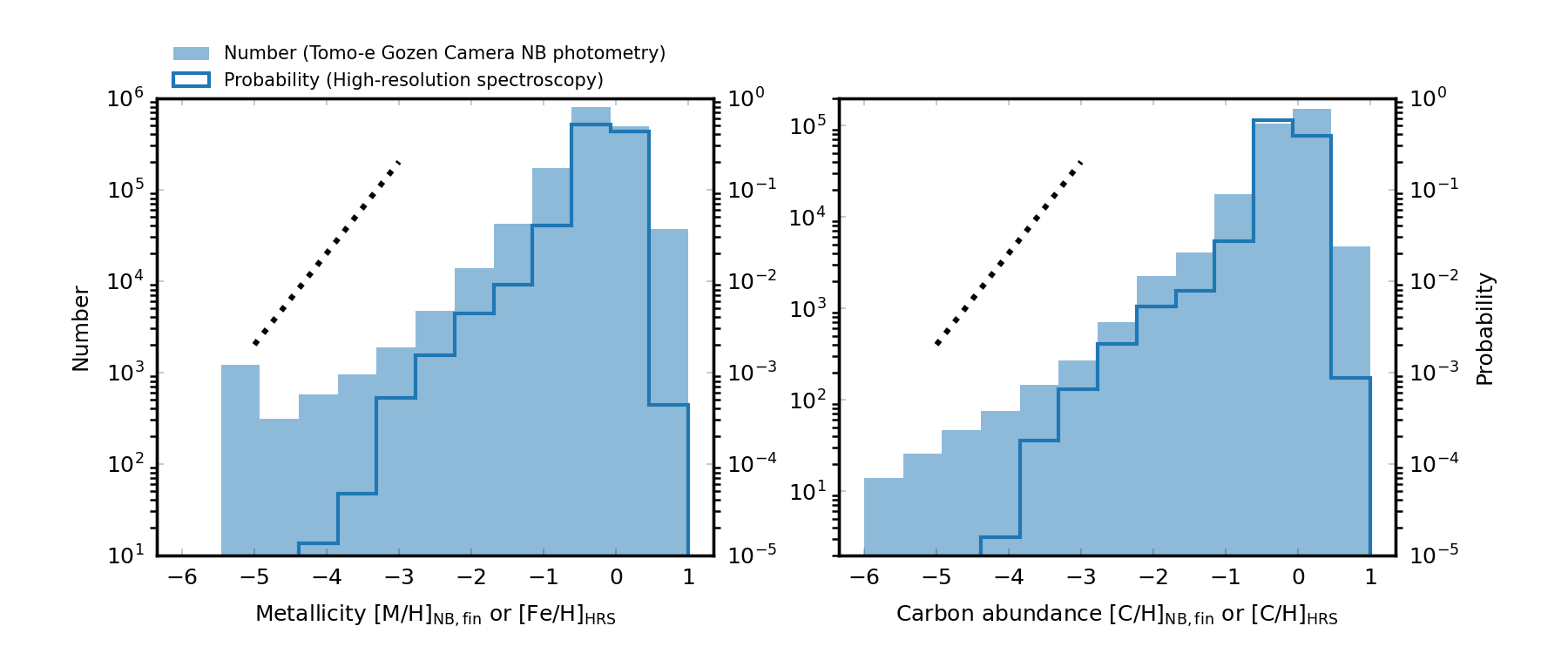}
 \end{center}
\caption{ Distribution functions of metallicity and carbon abundance for the targets (shaded region) and for stars with measurements from high-resolution spectra (open region). 
A dotted line corresponding to $\log_{10} N =$~[X/H]~$+ C$ is shown as a guide.
}\label{fig:MDF}
\end{figure*}

The distributions of metallicity and carbon abundance are shown in Figures~\ref{fig:MDF}. The distributions are compared with the probability distributions of stars with measurements from high-resolution spectroscopy. The distributions peak at similar values and have similar gradients down to \MH~$\sim-3$ and \CH~$\sim-3$. This would imply that our estimates of metallicity and carbon abundance are reliable down to \MH~$\sim-3$ and \CH~$\sim-3$. In contrast to the distributions obtained from high-resolution spectroscopic measurements, our distributions extend down to \MH~$\sim-5$ and \CH~$\sim-6$ with similar slopes, with an enhancement at \MH~$\sim-5$. This might indicate false estimates at \MH~$<-3$ or \CH~$<-3$, but some metal-poor stars with \MH~$<-3$ and \CH~$<-3$ could be included in our targets. 
We note that the enhancement at \MH~$\sim-5$ is likely to be a false positive because the metallicity lies at the edge of the model parameter range. 

\subsection{Comparison of the photometric and spectroscopic metallicities}

\begin{figure}[htbp]
  \begin{center}
    \includegraphics[clip,width=\columnwidth]{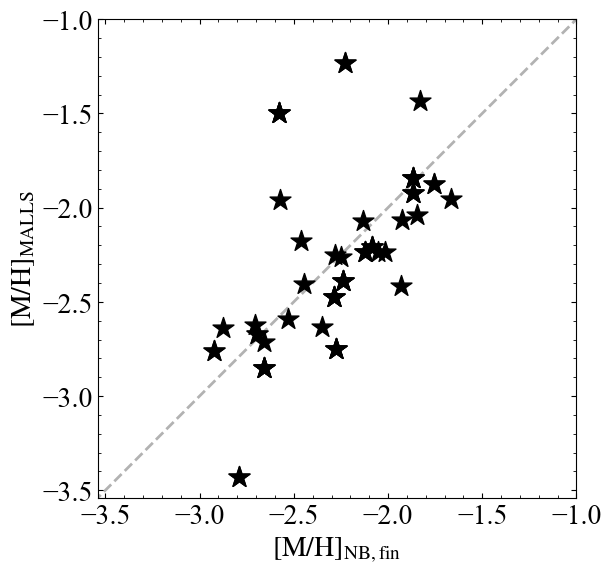}
  \end{center}
    \caption{Comparison between the metallicity estimated from NB photometry, \MHnbf, and that determined from medium-resolution spectroscopy, \FeHmalls.
   }
    \label{fig:TomoeMR}
\end{figure}

Thirty-two metal-poor star candidates identified using the Tomo-e Gozen camera were followed up with medium-resolution spectroscopy using Nayuta/MALLS. Their metallicities are determined by the method described in Section~\ref{sec:MALLSreduction}. Figure~\ref{fig:TomoeMR} demonstrates that the metallicities estimated from the NB photometry are mostly consistent with those determined from the medium-resolution spectra.
Among the 32 candidates, one object satisfies \FeHmalls~$\leq -3$; this object has the lowest metallicity in this sample, with \FeHmalls~$= -3.4$. In addition, 23 objects satisfy $-3 <$~\FeHmalls~$\leq -2$.

There are two outliers with \MHnbf~$< -2$ and \FeHmalls~$> -1.5$. One target with \FeHmalls~$= -1.2$ has relatively low \SN, \sNB~$= 0.06$. In contrast, the other target with \FeHmalls~$= -1.5$ has high \SN, \sNB~$= 0.02$, but meets the criterion $C^* > \sigma_{C}^*$. This indicates that the outliers can be excluded by applying more stringent quality cuts.

The target with \FeHmalls~$= -3.4$ has \MHnbf~$= -2.7$. The relatively high value of \MHnbf\ may be due to the target's modest \SN, \sNB~$= 0.04$. The metallicity of this object was independently measured to be \FeH~$=-3.4$ using high-resolution spectroscopy with the Subaru/High Dispersion Spectrograph (HDS). A detailed elemental-abundance analysis based on the high-resolution spectrum will be presented in a forthcoming paper (Okada et al., in preparation).

\section{Summary}\label{sec:summary}

We are conducting the Tomo-e Gozen Bright Metal-Poor Star Survey (TeMPS). The survey employs four NB filters, NB395, NB411, NB433, and NB656. We have covered $\gtrsim 22,000$~deg$^2$ in all four bands, with a total on-source integration time of $\sim 100$ hours. The median depths at \SN~$=20$ are $14.0$~mag in the $NB395$ band, $14.3$~mag in the $NB411$ band, $15.2$~mag in the $NB433$ band, and $13.0$~mag in the $NB656$ band. The depths correspond to approximately $G \sim 12.5$.

We estimate metallicities for $\sim1,700,000$~targets and identify $\sim16,000$ VMP star candidates.
The metallicity and carbon abundance estimates have typical standard deviations of $\lesssim 0.3$ dex and $\lesssim 0.4$ dex, respectively.

We also carried out medium-resolution spectroscopic follow-up observations with MALLS on the Nayuta telescope for metal-poor star candidates selected from the NB photometry. Of the 32 candidates, 24 were confirmed to have \FeHmalls~$< -2$, including one star with \FeHmalls~$< -3$. This demonstrates that the Tomo-e Gozen selection efficiently identifies genuinely metal-poor stars.

In two color–color diagrams, metal-poor stars form a distinct locus that is clearly separated from the population of stars with \FeH~$\sim0$. As shown in previous studies, the $NB395 - g - 1.5(g-i)$ and $NB395 - NB656 - 2(g-i)$ colors efficiently discriminate metal-poor stars from stars with solar metallicity. Interestingly, at a fixed $NB395 - NB411$, the typical separation corresponds to $\sim 0.4$~mag in the $NB395 - g - 1.5(g-i)$ color. This separation is larger than that obtained at a fixed $g-i$, which is commonly adopted for metal-poor star selection. We also show that the $g-i$ and $NB395 - NB433$ color combination distinguishes stars with different carbon abundances.

We compare our estimated metallicities with those reported in previous studies. In most comparisons, some targets with \MHnbf~$< -2$ are reported or suggested to have \MH~$> -2$ in other studies. This indicates that our selection is tuned for high completeness at the expense of purity. Such a choice is well matched to our strategy: we first select candidates using NB photometry and then confirm their metallicities with medium-resolution spectroscopy.
The inferred metallicity and carbon-abundance distribution functions exhibit slopes consistent with those derived from stars with measurements based on high-resolution spectra, down to \MH~$\sim -3$ and \CH~$\sim -3$. The apparent extensions toward \MH~$< -3$ and \CH~$< -3$ will be tested with future follow-up observations.

Through the Tomo-e Gozen NB photometric survey and Nayuta/MALLS medium-resolution follow-up, we have developed a two-stage strategy to efficiently identify bright metal-poor stars: photometric candidate selection followed by spectroscopic metallicity confirmation. The Tomo-e Gozen Bright Metal-Poor Star Survey will continue until the northern sky is fully covered, and Nayuta/MALLS follow-up will provide confirmed metal-poor stars as prime targets for high-resolution spectroscopic observations, including those with Subaru/HDS.

\begin{ack}
The spherical filter holder was manufactured at Advanced Technology Center, National Astronomical Observatory of Japan (2021-034 and 2022-029).
Data analyses were in part carried out on  PC cluster and GPU cluster at Center for Computational Astrophysics, National Astronomical Observatory of Japan.
This research has been supported in part by the Grant-in-Aid for Scientific Research of the JSPS (JP21H04499, JP22K03688, JP23H04894, JP24K00682, JP25K01046, JP25H00674, JP25H02196, JP25HP8001).
H.O. was supported by the NAOJ Overseas Visit Program for Young Researchers (FY2024) and a grant from the Hayakawa Satio Fund awarded by the Astronomical Society of Japan.
\end{ack}








\bibliographystyle{hapj} 
\bibliography{ms}

\end{document}